\documentclass[11pt]{article}
\usepackage{epsfig}
\begin{document}
\vspace*{-2cm}
\noindent
\begin{flushright}
% \hspace*{12.3cm}
UG--FT--128/01 \\
LC--TH--2001--067 \\
% \hspace*{12.3cm}
hep-ph/0102197 \\
% \hspace*{12.3cm}
February 2001 \\
\end{flushright}
\vspace{0.5cm}
\begin{center}
\begin{Large}
{\bf Probing top flavour-changing neutral couplings at TESLA}
\end{Large}

\vspace{0.5cm}
J. A. Aguilar--Saavedra \\
{\it Departamento de F\'{\i}sica Te\'{o}rica y del Cosmos,
Universidad de Granada \\
E-18071 Granada, Spain} \\
\vspace{0.2cm}

T. Riemann \\
{\it Deutsches Elektronen-Synchrotron DESY \\ 
Platanenallee 6, D-15738 Zeuthen, Germany} \\
\end{center}

\begin{abstract}
We present a comprehensive analysis of the sensitivity of the TESLA $e^+ e^-$
collider to top
flavour-changing neutral couplings to the $Z$ boson and photon. We study single
top production and top decay processes, and we consider the cases without beam
polarization, with only $e^-$ polarization and with $e^-$ and $e^+$
polarization. We show that the use of the latter substantially enhances the
sensitivity to discover or bound these vertices, and for some of the couplings
the expected LHC limits could be improved by factors $2-14$ for equal running
times.
%\vspace*{0.5cm} \noindent
%PACS:  12.15.Mm; 12.60.-i; 14.65.Ha; 14.70.-e
% 12.15.Mm Neutral currents
% 12.60.-i Models beyond the standard model
% 14.65.Ha Top quarks
% 14.70.-e Gauge bosons
\end{abstract}

\section{Introduction}
\label{sec:1}
It is generally believed that the top quark, because of its large mass, will be
a sensitive probe into physics beyond the Standard Model (SM) \cite{papiro1}. In
particular, its couplings to the gauge and Higgs bosons may show deviations with
respect to the SM predictions. In the SM the flavour-changing neutral (FCN)
couplings $Ztq$, with $q=u,c$, vanish at tree-level
by the GIM mechanism, and the $\gamma tq$ and $gtq$ ones are zero as a
consequence of  the unbroken ${\mathrm SU(3)}_c \times {\mathrm U(1)}_Q$
symmetry. The $Htq$ couplings also vanish due to
the existence of only one Higgs doublet.
These types of vertices can be generated at the
one-loop level, but they are very suppressed by the GIM mechanism, because the
masses of the charge $-1/3$ quarks in the loop are small compared to the
scale involved. 
The single top production branching may be estimated roughly by  
$\mathrm{Br}(Z \to tc) = 1.5 \times 10^{-13}$ \cite{ztc},
and 
the calculation of the branching ratios for top decays mediated
by these FCN operators yields the SM predictions
$\mathrm{Br}(t \to Zc) = 1.3 \times 10^{-13}$,
$\mathrm{Br}(t \to \gamma c) = 4.5 \times 10^{-13}$,
$\mathrm{Br}(t \to gc) = 3.5 \times 10^{-11}$ \cite{papiro12},
$\mathrm{Br}(t \to Hc) = 3.5 \times 10^{-14}$ \cite{papiro12b}
\footnote{We assume a Higgs mass $M_H=120$ GeV.},
 and smaller values for the up quark.
 However, in many simple SM extensions these rates can
be orders of magnitude larger. For instance, in models with exotic quarks
$\mathrm{Br}(t \to Zq)$ can be of order $10^{-3}-10^{-2}$ \cite{papiro13}.
Two Higgs doublet models allow for $\mathrm{Br}(t \to Zc) = 10^{-6}$,
$\mathrm{Br}(t \to \gamma c) = 10^{-7}$,
$\mathrm{Br}(t \to gc) = 10^{-4}$ \cite{papiro14}, and in $R$ 
parity-violating supersymmetric models one can have
$\mathrm{Br}(t \to Zc) = 10^{-4}$, $\mathrm{Br}(t \to \gamma c) = 10^{-5}$,
$\mathrm{Br}(t \to gc) = 10^{-3}$
\cite{papiro15}. Top FCN decays into a light Higgs boson and an up or charm
quark can also have similar or larger rates in models with exotic quarks
\cite{papiro27,papiro13}, with more than one Higgs doublet
\cite{papiro14,papiro28a} or with supersymmetry
\cite{papiro28}. Hence, top FCN couplings offer a good
place to search for new
physics, which may manifest if these vertices are observed in
future colliders. In addition, the study of FCN couplings provides
model-independent information
on the charged current couplings and the unitarity of the CKM matrix
\cite{papiro29}.
Here we will focus on FCN
interactions involving the top, a light charge $2/3$ quark $q$ and a neutral
gauge boson $V=Z,\gamma$.
At present the best limits on $Ztq$ couplings come from LEP 2,
$\mathrm{Br}(t \to Zq) \leq 0.07$ \cite{papiro25a,papiro25b}, and the best
limits on $\gamma tq$ couplings from Tevatron, $\mathrm{Br}(t \to \gamma q) \leq
0.032$
\cite{papiro26}. They are very weak but will improve in the next years, first
with Tevatron Run II, and later with the next generation of colliders.

The CERN LHC will be a top factory. With a $t \bar t$ production cross-section
of 830 pb, at its 100 fb$^{-1}$ luminosity phase it will produce
$8.3 \times 10^7$ top-antitop pairs per year. In addition, it will produce
$3 \times 10^7$ single tops plus antitops via other processes
\cite{papiro2,papiro3}. This makes LHC an excellent machine
for the investigation of the top quark properties. The search for FCN top
couplings can be carried out examining two different types of processes. On the
one hand, we can look for rare top decays $t \to Zq$ \cite{papiro4},
$t \to \gamma q$ \cite{papiro5}, $t \to gq$ \cite{papiro6} or
$t \to Hq$ \cite{papiro7} of the tops or antitops produced in the SM process
$gg,q \bar q \to t \bar t$. On the other hand, one can search for single top
production via an anomalous effective vertex: $Zt$ and $\gamma t$ production
\cite{papiro8}, the production of a top quark without or with a light
jet \cite{papiro9,papiro10}, and $Ht$ production \cite{papiro7}. In these cases
the top quark is assumed to decay in the SM dominant mode $t \to Wb$. One can
also search for like-sign $tt$ production \cite{papiro11} and other exotic
processes.

The TESLA $e^+ e^-$ collider with a centre of mass (CM) energy of
$\sqrt s = 500$ GeV has a tree-level
$t \bar t$ production cross-section of 0.52 pb, and produces only
$1.56 \times 10^5$ top-antitop pairs per year with its expected luminosity of
300 fb$^{-1}$. However, $e^+ e^-$
colliders are cleaner than hadron colliders. For instance, the signal to
background ratio $S/B$ for rare top decays can be 7 times larger in TESLA than
in LHC. But the sensitivity to rare top decays is given in the Gaussian
statistics limit by the ratio $S/\sqrt B$, and the larger LHC cross-sections
make difficult for TESLA to compete with it in the search for anomalous top
couplings.

In this paper we show that the use of beam polarization in TESLA substantially
enhances its sensitivity to discover or bound top anomalous FCN couplings and
allows to improve some of the expected LHC limits up to an order of magnitude
\cite{papiro0}.
We first study the single top production process $e^+ e^- \to t \bar q$,
mediated by $Ztq$ or $\gamma tq$ anomalous couplings
\cite{papiro16}. Then we study rare top decays in the
processes $e^+ e^- \to t \bar t$, with subsequent decay $\bar t \to V \bar q$.
In all cases we take into account the charge conjugate processes as well:
we sum $t \bar q + \bar t q$ production, and we consider $\bar t \to V \bar q$
or $t \to Vq$. Single top production and top decay processes are complementary.
Single top production is more sensitive to top anomalous couplings but top
decays can help to disentangle the type of anomalous coupling involved ($Ztq$
or $\gamma tq$) if a positive signal is discovered.

We consider the planned CM energies of 500 and 800 GeV, and for both we analyse
the cases: ({\em i\/}) without beam polarization, ({\em ii\/}) with $80\%$
$e^-$ polarization, and ({\em iii\/}) with $80\%$ $e^-$, $45\%$ $e^+$
polarization. The paper is organized as follows. In Section~\ref{sec:2} we
describe the procedure used to compute the signals and backgrounds and to
obtain the limits. In Section~\ref{sec:3} we analyse single top production. In
Sections~\ref{sec:4} and \ref{sec:5} we consider top decays $t \to \gamma q$
and $t \to Zq$, respectively. In Section~\ref{sec:6} we summarize the results
and draw our conclusions.

\newpage
\section{Generation of signals and backgrounds}
\label{sec:2}
In order to describe the FCN couplings among the top, a light quark $q$ and a
$Z$ boson or a photon $A$ we use the Lagrangian
\begin{eqnarray}
-{\mathcal L} & = & \frac{g_W}{2 c_W} \, X_{tq} \, \bar t \gamma_\mu  
(x_{tq}^L P_L + x_{tq}^R P_R) q Z^\mu 
+ \frac{g_W}{2 c_W} \, \kappa_{tq}\, \bar t (\kappa_{tq}^{v}- \kappa_{tq}^{a}
\gamma_5) \frac{i \sigma_{\mu \nu} q^\nu}{m_t} q Z^\mu  \nonumber \\
& &  + e \, \lambda_{tq}\, \bar t (\lambda_{tq}^{v}- \lambda_{tq}^{a} \gamma_5)
\frac{i \sigma_{\mu \nu} q^\nu}{m_t} q A^\mu  \,, \label{ec:1}
\end{eqnarray}
where $P_{R,L}=(1 \pm \gamma_5)/2$. The chirality-dependent parts are normalized
to $(x_{tq}^L)^2+(x_{tq}^R)^2=1$, $(\kappa_{tq}^{v})^2+(\kappa_{tq}^{a})^2=1$,
$(\lambda_{tq}^{v})^2+(\lambda_{tq}^{a})^2=1$. This effective Lagrangian
contains $\gamma_\mu$ terms of dimension 4 and $\sigma_{\mu \nu}$ terms of
dimension 5. The couplings are constants corresponding to the first terms in
the expansion in momenta. The $\sigma_{\mu \nu}$ terms are the only ones allowed
by the unbroken gauge symmetry, ${\mathrm SU(3)}_c \times {\mathrm U(1)}_Q$. Due
to their extra momentum factor they grow with the energy and make large
colliders the best places to measure them.

\begin{figure}[htb]
\begin{center}
\mbox{\epsfig{file=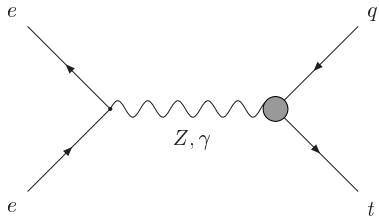,width=5cm}}
\end{center}
\caption{Feynman diagrams for $e^+ e^- \to t\bar q$ via $Ztq$ or $\gamma tq$
FCN couplings. The top quark is off-shell and has the SM decay.
\label{fig:feyn1} }
\end{figure}

For single top production we study the process $e^+ e^- \to t \bar q$ mediated
by $Ztq$ or $\gamma tq$ anomalous couplings (see Fig.~\ref{fig:feyn1}). We will
only take one anomalous coupling different from zero at the same time. However,
if a positive signal is discovered, it may be difficult to distinguish only from
this process whether the anomalous coupling involves the $Z$ boson, the photon
or both. On the other hand, in principle it could be possible to have a
fine-tuned cancellation between $Z$ and $\gamma$ contributions that led to a
suppression of this signal.

\begin{figure}[htb]
\begin{center}
\mbox{\epsfig{file=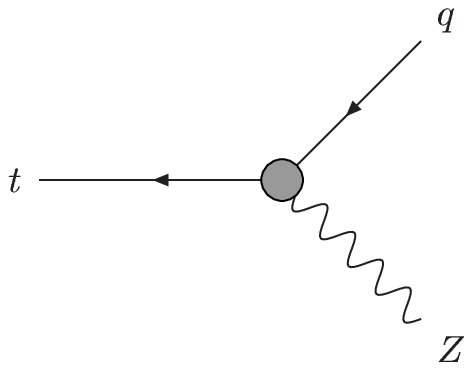,width=3.5cm}} ~~~
\mbox{\epsfig{file=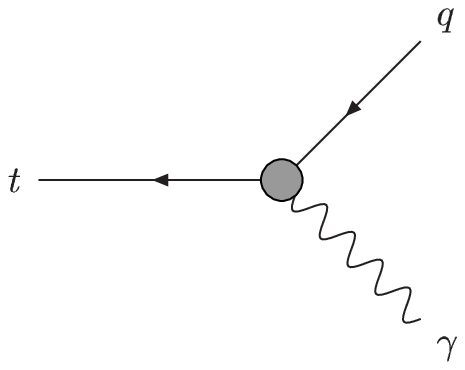,width=3.5cm}}
\end{center}
\caption{Feynman diagrams for FCN antitop decay via $Ztq$ or $\gamma tq$
FCN couplings.
\label{fig:feyn2} }
\end{figure}

For top decays we study the SM process $e^+ e^- \to t \bar t$, followed by
antitop decay mediated by an anomalous $Ztq$ or $\gamma tq$ coupling (see
Fig.~\ref{fig:feyn2}). This gives the signals $t \bar qZ$ and $t \bar q\gamma$,
and the observation of the final state distinguishes $Ztq$ and $\gamma tq$
couplings. In the $t\bar q$, $t \bar qZ$ and $t \bar q\gamma$ signals the
top is assumed to decay via $t \to W^+b \to l^+\nu b$, with $l=e,\mu$. For the
$t \bar qZ$ signal we only consider the $Z$ boson decays to electrons and muons.

For the $t \bar q$ signal we calculate the matrix element
$e^+ e^- \to t \bar q \to W^+ b \bar q \to l^+ \nu b \bar q$. For
the $t \bar qZ$ and $t \bar q\gamma$ signals we calculate
$e^+ e^- \to t \bar t \to W^+ b \bar q Z \to l^+ \nu b \bar q l'^+ l'^-$ and
$e^+ e^- \to t \bar t \to W^+ b \bar q \gamma \to l^+ \nu b \bar q \gamma$,
respectively. These matrix elements are evaluated using HELAS \cite{papiro18}
and introducing a new HELAS-like subroutine {\tt IOV2XX} to compute the
non-renormalizable $\sigma_{\mu \nu}$ vertex. This new routine has been checked
by hand. In all cases we sum the contribution of the charge conjugate processes.
For the $t \bar qV$ signals there is an additional contribution from $t \bar q$
production plus radiative emission of a $Z$ boson or a photon. This correction
is suppressed because it does not have the enhancement due to the $\bar t$
on-shell, and is even smaller after the kinematical cuts for the signal
reconstruction.

The background for the $t \bar q$ signal is given by $W^+ q \bar q'$ production
with $W^+$ decay to electrons and muons. The leading contribution to this
process is $W^+ W^-$ production with $W^-$ hadronic decay, but it is crucial for
the correct evaluation of the background after kinematical cuts to take into
account the 7 interfering Feynman diagrams for $e^+ e^- \to W^+ q \bar q'$.
Taking all the interfering diagrams for $e^+ e^- \to l^+ \nu q \bar q'$ into
account does not make any appreciable change in the cross-section. The
backgrounds for the $t \bar qZ$ and $t \bar q\gamma$ signals are analogous,
$W^+ q \bar q'Z$ and $W^+ q \bar q'\gamma$, with 46 and 44 diagrams,
respectively. These three backgrounds are evaluated using MadGraph
\cite{papiro19} and modifying the code to include the $W^+$ decay.

To simulate the calorimeter energy resolution we perform a Gaussian smearing of
the charged lepton ($l$), photon ($\gamma$) and jet ($j$) energies using a
realistic calorimeter resolution \cite{papiro20} of
\begin{equation}
\frac{\Delta E^{l,\gamma}}{E^{l,\gamma}} = \frac{10\%}{\sqrt{E^{l,\gamma}}}
\oplus 1\% \,, ~~~~
\frac{\Delta E^{j}}{E^{j}} = \frac{50\%}{\sqrt{E^j}} \oplus 4\% \,,
\end{equation}
where the energies are in GeV and the two terms are added in quadrature. For
simplicity we assume that the energy smearing for muons is the same as for
electrons. Note that more optimistic resolutions would improve our results. We
then apply detector cuts on transverse momenta $p_T$ and pseudorapidities $\eta$
\begin{equation}
p_T \geq 10 ~\mathrm{GeV} ~,~~ |\eta| \leq 2.5 \,.
\end{equation}
The cut on pseudorapidity corresponds to a polar angle $10^\circ \leq \theta
\leq 170^\circ$. We reject the events in which the jets and/or leptons are not
isolated, requiring that the distances in $(\eta,\phi)$ space $\Delta R$
satisfy $\Delta R \geq 0.4$. We do not require specific trigger conditions,
and we assume that the presence of high $p_T$ charged leptons will suffice.

After signal and background reconstruction, which will be analysed in detail for
each of the processes discussed, we require a $b$ tag on the jet associated to
the decay of the top quark to reduce the backgrounds. We require the $b$ tagged
jet to have $|\eta_b| \leq 2$ (polar angle $15^\circ \leq \theta_b \leq
165^\circ$) and energy $E_b \geq 45$ GeV. We assume a $b$ tagging efficiency of
$60\%$, and mistagging rates of $5\%$ for charm and $0.5\%$ for lighter quarks
\cite{papiro21}. These are average numbers appropriate for the $E_b$
kinematical distributions we will obtain later.

After kinematical cuts, for each of the cases studied we obtain two types of
limits on the anomalous coupling parameters $X_{tq}$, $\kappa_{tq}$,
$\lambda_{tq}$. Below we outline the procedure used.
The correct statistical treatment of
signals and backgrounds is specially necessary in our study since the
backgrounds are very small, sometimes much less than one event even for high
integrated luminosities.

Assuming that no signal is observed after the experiment is done, {\em i.e.}
the number of observed events $n_0$ equals the expected background $n_b$, we
derive $95\%$ confidence level (CL) upper bounds on the number of events
expected $n_s$. We use the Feldman-Cousins construction for the confidence
intervals of a Poisson variable \cite{papiro22} evaluated with the PCI package
\cite{papiro23}.

On the other hand, we can obtain the smallest value of $n_s$ such that a
positive signal is expected to be observed with $3\,\sigma$ significance,
assuming that the number of observed events for $3\,\sigma$ `evidence' $n_e$
equals $n_s+n_b$. For a large number of background events, the Poisson
probability distribution can be approximated by a Gaussian of mean $n_b$ and
standard deviation $\sqrt n_b$. The requirement of $3\,\sigma$ significance is
then simply $n_s/\sqrt n_b \geq 3$. However, this is seldom the case for our
study, where the backgrounds are very small. In such case, we use the estimator
based on the $\mathcal{P}$ number (see for example \cite{papiro24}). The number
$\mathcal{P}(n)$ is defined as the probability of the background to fluctuate
and give $n$ or more observed events. $n_e$ is then defined as the smallest
value of $n$ such that $1-\mathcal{P}(n) \geq 99.73\%$, corresponding to three
Gaussian standard deviations.

Another possible estimator for the evidence of a signal can be built using
Feldman-Cousins intervals. For a fixed value of $n_b$, we define the number of
observed events for $3\,\sigma$ evidence $n_e$ as the smallest value of the
number of observed events $n$ such that the $99.73\%$ CL Feldman-Cousins
intervals do not contain zero. These two estimators for the evidence of a
signal can be shown to be equivalent, and for $n_b \geq 9$ both give similar
results to the Gaussian approximation $n_s/\sqrt n_b \geq 3$.

In our analysis we find that usually $3\,\sigma$ evidence limits are numerically
larger than $95\%$ upper limits, but this is not always the case. For very small
backgrounds $3\,\sigma$ limits are smaller, what means that the potential to
discover a new signal is better than the ability to set upper bounds on it if
nothing is seen. This behaviour can exhibit fluctuations resulting
from the discreteness of Poisson statistics, but in general and comparing with
LHC, the TESLA discovery potential is better than the potential to set upper
limits.

\section{Single top production $e^+ e^- \to t \bar q$}
\label{sec:3}
The process $e^+ e^- \to t \bar q$ gives better limits on the top anomalous
couplings than top decays. However, it has the disadvantage that the final state
$l^+ \nu bj$ does not distinguish the type of coupling involved.  The background
is $W^+ jj$ production with $W^+ \to l^+ \nu$ and a jet misidentified as a $b$.

We take only one type of FCN couplings different from zero at the same time,
and we evaluate three signals: ({\em i\/}) with $Ztq$ $\gamma_\mu$, ({\em ii\/})
with $Ztq$ $\sigma_{\mu \nu}$, and ({\em iii\/}) with $\gamma tq$ couplings.
Their cross-sections depend slightly on the chirality of the anomalous
couplings. The chirality-dependent parts can be written as $(v-a \,\gamma_5)$,
with $v^2+a^2=1$. For definiteness, we set $v=1$, $a=0$ in our evaluations. The
results are the same setting $v=0$ and $a=1$. For a CM energy of 500 GeV and the
three polarization options discussed the cross-section for $\gamma_\mu$
couplings differs $\mp 1\%$ setting $v=\pm a = 1/\sqrt 2$, and for
$\sigma_{\mu \nu}$ couplings it differs $\pm 1.2\%$.

\begin{figure}[htb]
\begin{center}
\mbox{\epsfig{file=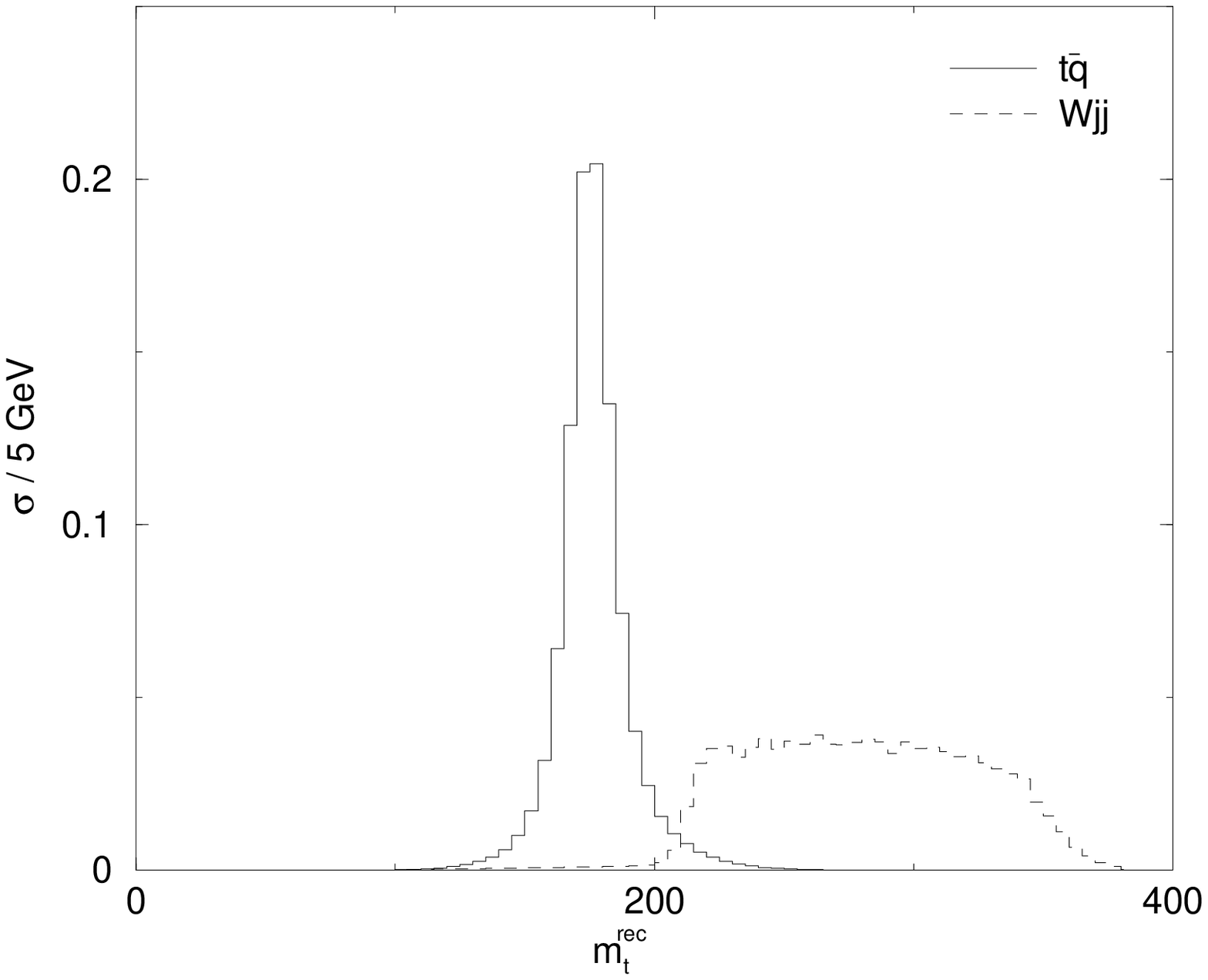,width=9cm}}
\end{center}
\caption{Reconstructed top mass $m_t^\mathrm{rec}$ distribution before
kinematical cuts for the three $t \bar q$ signals and $W^+ jj$ background at a
CM energy of 500 GeV, without beam polarization. The cross-sections are
normalized to unity.
\label{fig:tc-mt}}
\end{figure}

\begin{figure}[htb]
\begin{center}
\mbox{\epsfig{file=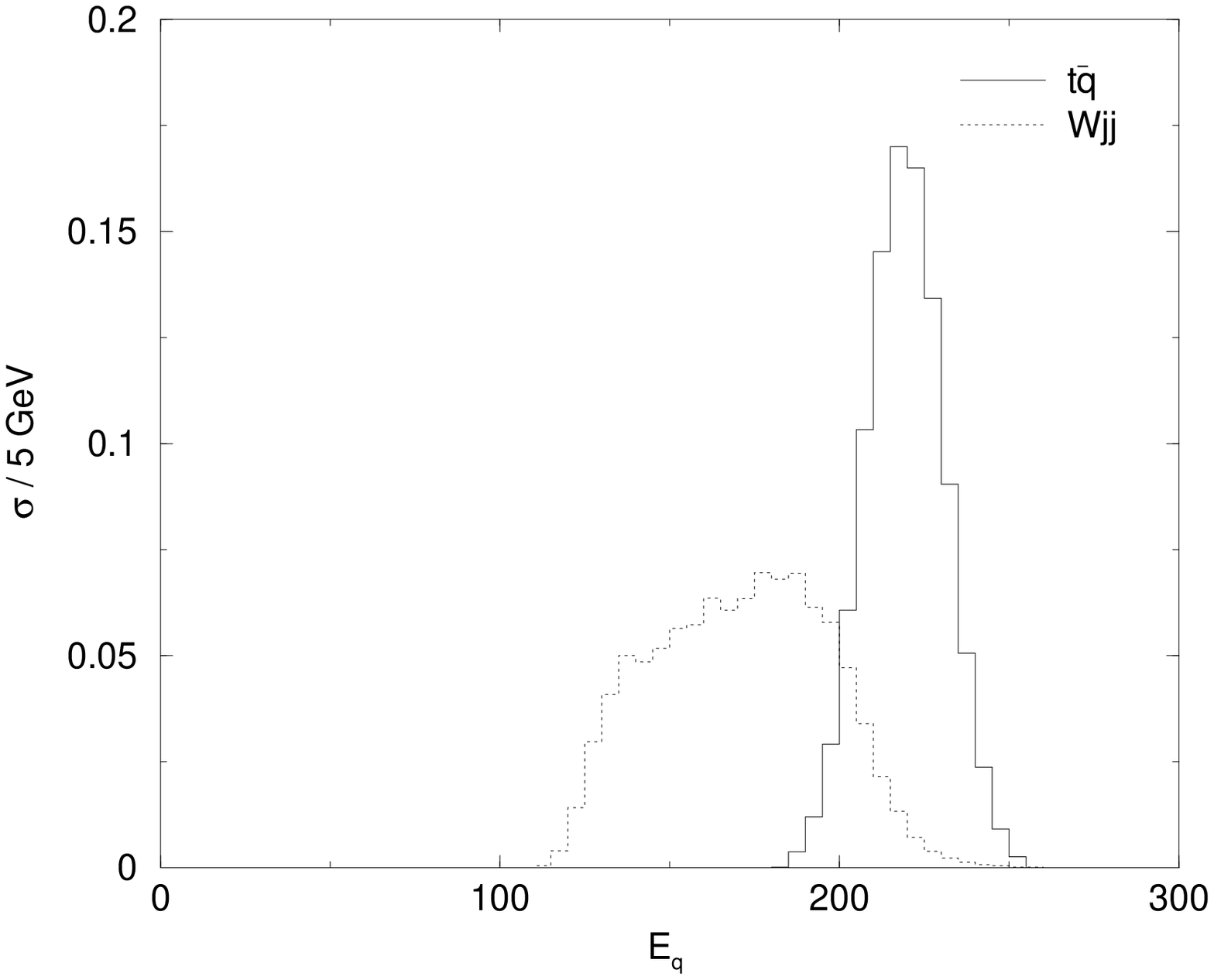,width=9cm}}
\end{center}
\caption{$E_q$ distribution before kinematical cuts for the three $t \bar q$
signals and $W^+ jj$ background at a CM energy of 500 GeV, without beam
polarization. The cross-sections are normalized to unity.
\label{fig:tc-ec}}
\end{figure}

\begin{figure}[htb]
\begin{center}
\mbox{\epsfig{file=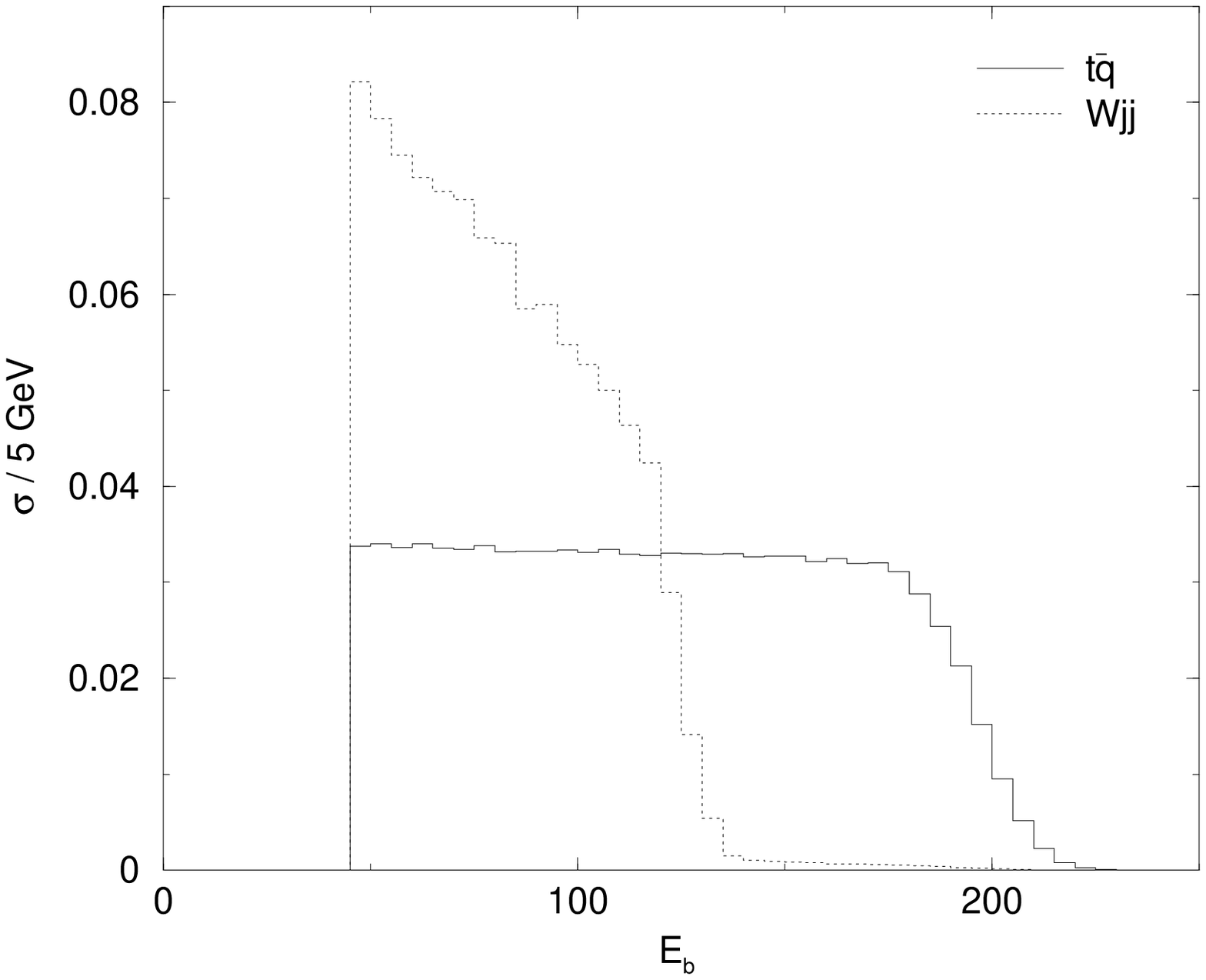,width=9cm}}
\end{center}
\caption{$E_b$ distribution before kinematical cuts for the three $t \bar q$
signals and $W^+ jj$ background at a CM energy of 500 GeV, without beam
polarization. The cross-sections are normalized to unity.
\label{fig:tc-eb}}
\end{figure}

The signals are reconstructed as follows. The neutrino momentum $p_\nu$ can be
identified with the missing momentum of the event. The longitudinal missing
momentum can also be used, and $p_\nu$ is reconstructed without any ambiguity.
The $W^+$ momentum is then the sum of the momenta of the charged lepton and the
neutrino. In the case of $t \bar q$ production, the invariant mass of the $W^+$
and one of the jets, $m_t^\mathrm{rec}$, is consistent with the top mass,
and the other jet has an energy $E_q$ around
$E_q^0 \equiv (s-m_t^2)/(2 \sqrt s)$. Of the
two possible pairings, we choose the one minimizing
$(m_t^\mathrm{rec}-m_t)^2+(E_q-E_q^0)^2$ and require a $b$ tag on the jet
associated to the top quark. The kinematical distributions for
$m_t^\mathrm{rec}$ and $E_q$ for the signals and the background at a CM
energy of 500 GeV without beam polarization are plotted in
Figs.~\ref{fig:tc-mt} and \ref{fig:tc-ec}. The $E_b$ distribution is shown in
Fig.~\ref{fig:tc-eb}.

\begin{figure}[htb]
\begin{center}
\mbox{\epsfig{file=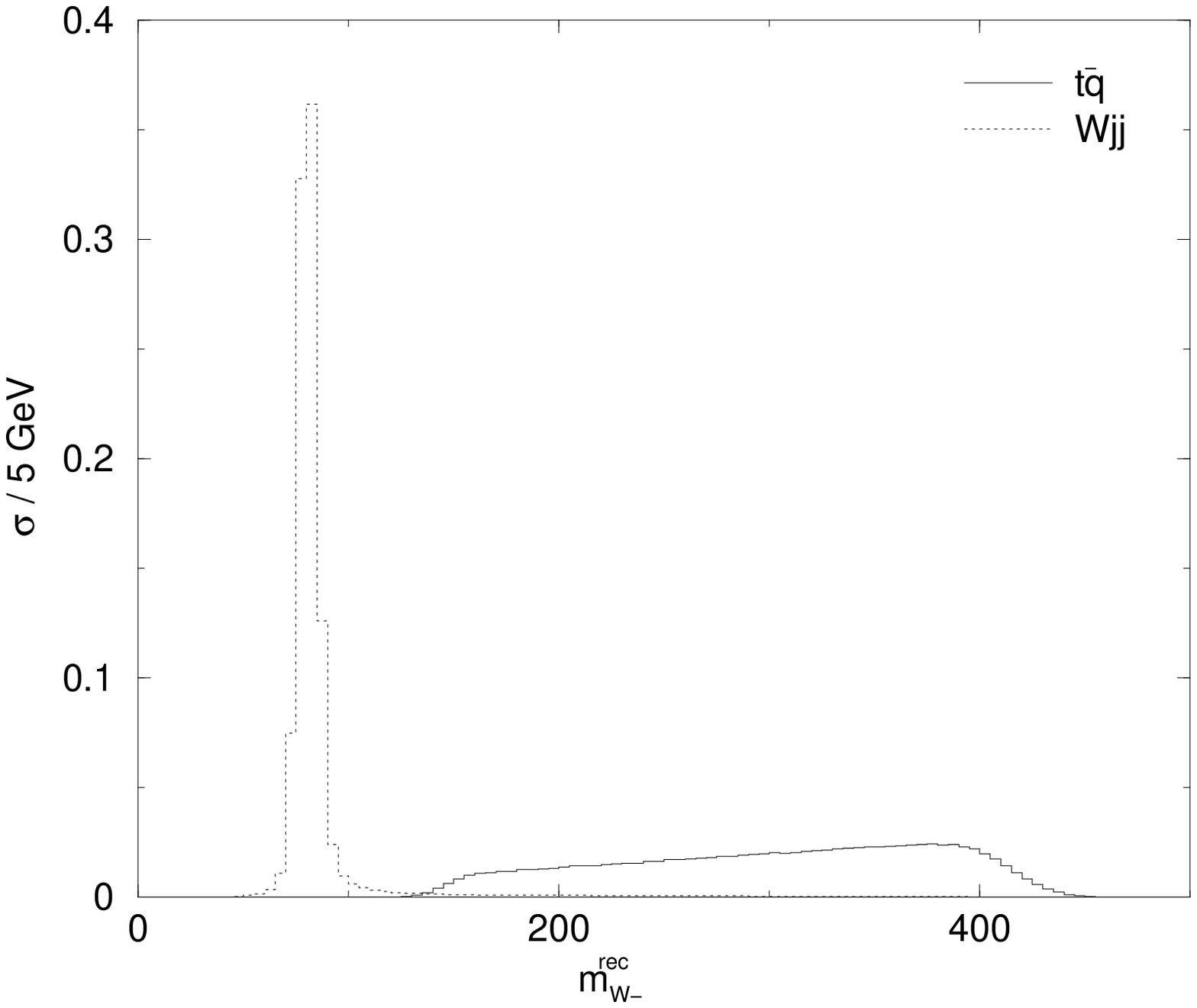,width=9cm}}
\end{center}
\caption{Reconstructed $W^-$ mass $M_{W^-}^\mathrm{rec}$ distribution before
kinematical cuts for the three $t \bar q$ signals and $W^+ jj$ background at a
CM energy of 500 GeV, without beam polarization. The cross-sections are
normalized to unity.
\label{fig:tc-mw2}}
\end{figure}

\begin{figure}[htb]
\begin{center}
\mbox{\epsfig{file=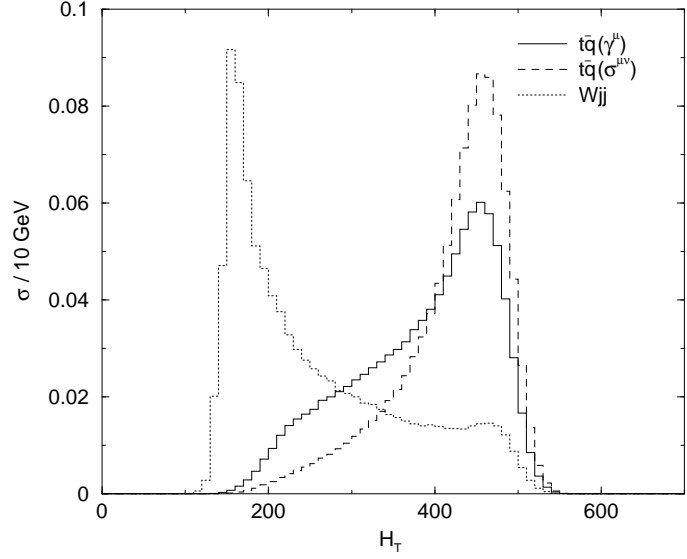,width=9cm}}
\end{center}
\caption{Total transverse energy $H_T$ distribution before kinematical cuts for
the three $t \bar q$ signals and $W^+ jj$ background at a CM energy of 500 GeV,
without beam polarization. The cross-sections are normalized to unity.
\label{fig:tc-ht}}
\end{figure}

\begin{figure}[htb]
\begin{center}
\mbox{\epsfig{file=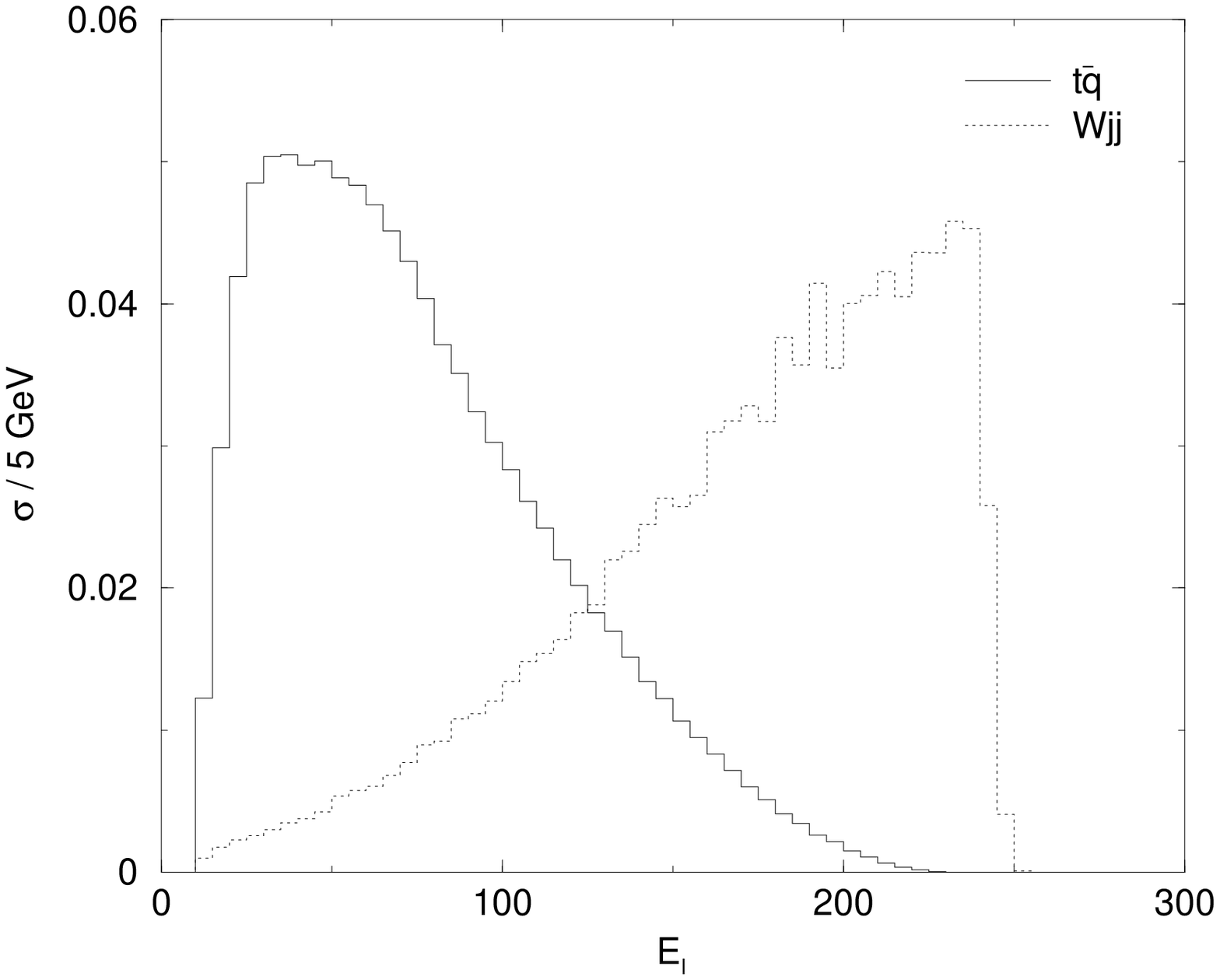,width=9cm}}
\end{center}
\caption{Charged lepton energy $E_l$ distribution before kinematical cuts for
the three $t \bar q$ signals and $W^+ jj$ background at a CM energy of 500 GeV,
without beam polarization. The cross-sections are normalized to unity.
\label{fig:tc-el}}
\end{figure}

Another interesting variable to distinguish the signals from the background
is the two-jet invariant mass $M_{W^-}^\mathrm{rec}$. The $W^+jj$ background is
dominated by $W^+ W^-$ production with $W^- \to jj$, and the
$M_{W^-}^\mathrm{rec}$ distribution peaks around $M_W$, as can be clearly seen
in Fig.~\ref{fig:tc-mw2}. A veto cut on $M_{W^-}^\mathrm{rec}$ can eliminate a
large fraction of the background but makes compulsory to calculate correctly the
cross-section to include all the diagrams for $e^+ e^- \to W^+ q \bar q'$. Also
of interest are the total transverse energy $H_T$ and the charged lepton energy
$E_l$ in Figs.~\ref{fig:tc-ht} and \ref{fig:tc-el}. The kinematical
distributions with polarized beams are very similar except the $H_T$
distribution. In this case polarization decreases the peak of the background
around $H_T=200$.

To enhance the signal significance we perform kinematical cuts on these
variables. However, we find that the veto cut on $M_{W^-}^\mathrm{rec}$ is
unnecessary in single top production since the requirement $E_b>45$ GeV and the
kinematical cut on $m_t^\mathrm{rec}$ practically eliminate the peak in the
$M_{W^-}^\mathrm{rec}$
distribution. A cut on $E_q$ is unnecessary because this
variable is kinematically related to $m_t^\mathrm{rec}$, and we prefer to apply
a cut on $m_t^\mathrm{rec}$ to show the presence of a top quark in the signal.
For simplicity, we choose to apply the same cuts for the three
signals and the three polarization options, but different for CM energies of 500
and 800 GeV. We choose the cuts trying to maintain the independence of the
cross-section on the chirality of the coupling. Obviously, our results could be
improved modifying the cuts for each type of coupling and each polarization
option. We now discuss the results for 500 GeV and 800 GeV in turn.

\subsection{Limits at 500 GeV}

The kinematical cuts for 500 GeV are collected in Table~\ref{tab:tc1}, and the
cross-sections before and after cuts for the three polarization options in
Table~\ref{tab:tc2}. We normalize the signals to $X_{tq} = 0.06$,
$\kappa_{tq} = 0.02$, $\lambda_{tq} = 0.02$ and sum the charge conjugate
processes. For the chiralities with $v=\pm a$ the cross-section after cuts
differs $\pm 6.8\%$ for $\gamma_\mu$ couplings and $\mp 4.6\%$ for
$\sigma_{\mu \nu}$ couplings.

\begin{table}[htb]
\begin{center}
\begin{tabular}{cc}
Variable & 500 GeV cut \\
$m_t^\mathrm{rec}$ & 160--190 \\
$H_T$ & $>220$ \\
$E_l$ & $<160$ 
\end{tabular}
\caption{Kinematical cuts for the three $t \bar q$ signals and the three
polarization options at a CM energy of 500 GeV. The masses and the energies are
in GeV.
\label{tab:tc1}}
\end{center}
\end{table}

\begin{table}[htb]
\begin{center}
\begin{tabular}{ccccccc}
& \multicolumn{2}{c}{No pol.} & \multicolumn{2}{c}{Pol. $e^-$}
& \multicolumn{2}{c}{Pol. $e^-$ $e^+$} \\
& before & after & before & after & before & after \\[-0.2cm]
& cuts & cuts & cuts & cuts & cuts & cuts \\
$t \bar q+\bar tq$ $(Z,\gamma_\mu)$
           & 0.183 & 0.137 & 0.162 & 0.121 & 0.215 & 0.161 \\
$t \bar q+\bar tq$ $(Z,\sigma_{\mu \nu})$
           & 0.199 & 0.153 & 0.176 & 0.135 & 0.234 & 0.179 \\
$t \bar q+\bar tq$ $(\gamma)$
           & 0.375 & 0.288 & 0.375 & 0.287 & 0.510 & 0.391 \\
$W^\pm jj$ & 19.5  & 0.0734 & 4.06  & 0.0154 & 2.40 & 0.0092
\end{tabular}
\caption{Cross-sections (in fb) before and after the kinematical cuts in
Table~\ref{tab:tc1} for the three $t \bar q$ signals and their background at a
CM energy of 500 GeV, for the three polarization options. We include $b$ tagging
efficiencies and use $X_{tq} = 0.06$, $\kappa_{tq} = 0.02$,
$\lambda_{tq} = 0.02$.
\label{tab:tc2}}
\end{center}
\end{table}

We notice in Table~\ref{tab:tc2} the usefulness of polarization: $e^-$
polarization decreases the background by a factor of 5, without affecting too
much the signals. $e^+$ polarization further decreases the background and even
increases the cross-section of the signals with respect to the values without
polarization.

We express the limits on the anomalous couplings in terms of top decay branching
ratios, using $\Gamma_t = 1.56$ GeV.  As explained in the previous Section, we
obtain $95\%$ CL upper limits for the case that nothing is observed and
$3\,\sigma$ discovery limits. Since the number of background events is small,
the limits do not scale with the luminosity $L$ as $1/\sqrt L$. In
Table~\ref{tab:tc3} we quote limits for a reference integrated luminosity of 100
fb$^{-1}$ for comparison with other processes, and in Table~\ref{tab:tc4} for
300 fb$^{-1}$, corresponding to one year of operation with the expected
luminosity.

\begin{table}[htb]
\begin{center}
\begin{small}
\begin{tabular}{lcccccc}
& \multicolumn{2}{c}{No pol.} & \multicolumn{2}{c}{Pol. $e^-$}
& \multicolumn{2}{c}{Pol. $e^-$ $e^+$} \\
& $95\%$ & $3\,\sigma$ & $95\%$ & $3\,\sigma$ & $95\%$ & $3\,\sigma$ \\
$\mathrm{Br}(t \to Zq)$ $(\gamma_\mu)$ &
  $7.9 \times 10^{-4}$ & $1.2 \times 10^{-3}$ &
  $7.1 \times 10^{-4}$ & $7.5 \times 10^{-4}$ & 
  $4.4 \times 10^{-4}$ & $4.2 \times 10^{-4}$ \\
$\mathrm{Br}(t \to Zq)$ $(\sigma_{\mu \nu})$ &
  $6.3 \times 10^{-5}$ & $9.4 \times 10^{-5}$ &
  $5.7 \times 10^{-5}$ & $6.0 \times 10^{-5}$ &
  $3.5 \times 10^{-5}$ & $3.4 \times 10^{-5}$ \\
$\mathrm{Br}(t \to \gamma q)$ &
  $3.9 \times 10^{-5}$ & $5.9 \times 10^{-5}$ &
  $3.2 \times 10^{-5}$ & $3.3 \times 10^{-5}$ &
  $1.9 \times 10^{-5}$ & $1.8 \times 10^{-5}$
\end{tabular}
\end{small}
\caption{Limits on top FCN decays obtained from single top production at 500 GeV
with a reference luminosity of 100 fb$^{-1}$ for the three polarization options.
\label{tab:tc3}}
\end{center}
\end{table}

\begin{table}[htb]
\begin{center}
\begin{small}
\begin{tabular}{lcccccc}
& \multicolumn{2}{c}{No pol.} & \multicolumn{2}{c}{Pol. $e^-$}
& \multicolumn{2}{c}{Pol. $e^-$ $e^+$} \\
& $95\%$ & $3\,\sigma$ & $95\%$ & $3\,\sigma$ & $95\%$ & $3\,\sigma$ \\
$\mathrm{Br}(t \to Zq)$ $(\gamma_\mu)$ &
  $4.4 \times 10^{-4}$ & $6.1 \times 10^{-4}$ &
  $3.1 \times 10^{-4}$ & $3.9 \times 10^{-4}$ &
  $1.9 \times 10^{-4}$ & $2.2 \times 10^{-4}$ \\
$\mathrm{Br}(t \to Zq)$ $(\sigma_{\mu \nu})$ &
  $3.5 \times 10^{-5}$ & $4.8 \times 10^{-5}$ &
  $2.4 \times 10^{-5}$ & $3.1 \times 10^{-5}$ &
  $1.5 \times 10^{-5}$ & $1.7 \times 10^{-5}$ \\
$\mathrm{Br}(t \to \gamma q)$ &
  $2.2 \times 10^{-5}$ & $3.0 \times 10^{-5}$ &
  $1.3 \times 10^{-5}$ & $1.7 \times 10^{-5}$ &
  $8.2 \times 10^{-6}$ & $9.3 \times 10^{-6}$
\end{tabular}
\end{small}
\caption{Limits on top FCN decays obtained from single top production at 500
GeV with a luminosity of 300 fb$^{-1}$ for the three polarization options.
\label{tab:tc4}}
\end{center}
\end{table}

\subsection{Limits at 800 GeV}

We write the kinematical cuts for 800 GeV in Table~\ref{tab:tc5}, and the
cross-sections before and after cuts in
in Table \ref{tab:tc6}. We normalize the signals to $X_{tq} = 0.06$,
$\kappa_{tq} = 0.02$, $\lambda_{tq} = 0.02$ and sum the charge conjugate
processes.
The signal cross-sections with non-renormalizable  couplings do not decrease
going from 500 to 800 GeV, whereas the background decreases by a factor of 2.3.
This improves the sensitivity for $\sigma_{\mu \nu}$ couplings with respect to
500 GeV. Unfortunately, the signal with $\gamma_\mu$ couplings also decreases,
and thus the results are worse in this case. Again we observe the usefulness
of polarization: using only $e^-$ polarization reduces the background 5 times 
and using $e^+$ polarization as well reduces it 8 times. In Table~\ref{tab:tc7}
we gather the limits for a reference integrated luminosity of 100 fb$^{-1}$ and
in Table~\ref{tab:tc8} for 500 fb$^{-1}$, collected in one year with the
expected luminosity.

\begin{table}[htb]
\begin{center}
\begin{tabular}{ccc}
Variable & 800 GeV cut \\
$m_t^\mathrm{rec}$ & 160--190 \\
$H_T$ & $>450$ \\
$E_l$ & $<300$
\end{tabular}
\caption{Kinematical cuts for the three $t \bar q$ signals and the three
polarization options at a CM energy of 800 GeV. The masses and the energies are
in GeV.
\label{tab:tc5}}
\end{center}
\end{table}

\begin{table}[htb]
\begin{center}
\begin{tabular}{lcccccc}
& \multicolumn{2}{c}{No pol.} & \multicolumn{2}{c}{Pol. $e^-$}
& \multicolumn{2}{c}{Pol. $e^-$ $e^+$} \\
& before & after & before & after & before & after \\[-0.2cm]
& cuts & cuts & cuts & cuts & cuts & cuts \\
$t \bar q+\bar tq$ $(Z,\gamma_\mu)$
           & 0.0776 & 0.0498 & 0.0684 & 0.0440 & 0.0912 & 0.0586 \\
$t \bar q+\bar tq$ $(Z,\sigma_{\mu \nu})$
           & 0.198 & 0.149 & 0.175 & 0.132 & 0.233 & 0.175 \\
$t \bar q+\bar tq$ $(\gamma)$
           & 0.389 & 0.293 & 0.389 & 0.293 & 0.528 & 0.398 \\
$W^\pm jj$ & 8.45   & 0.0125 & 1.75   & 0.0028 & 1.03   & 0.0018
\end{tabular}
\caption{Cross-sections (in fb) before and after the kinematical cuts in
Table~\ref{tab:tc5} for the three $t \bar q$ signals and their background at a
CM energy of 800 GeV, for the three polarization options. We include $b$ tagging
efficiencies and use $X_{tq} = 0.06$, $\kappa_{tq} = 0.02$,
$\lambda_{tq} = 0.02$.
\label{tab:tc6}}
\end{center}
\end{table}

\begin{table}[htb]
\begin{center}
\begin{small}
\begin{tabular}{lcccccc}
& \multicolumn{2}{c}{No pol.} & \multicolumn{2}{c}{Pol. $e^-$}
& \multicolumn{2}{c}{Pol. $e^-$ $e^+$} \\
& $95\%$ & $3\,\sigma$ & $95\%$ & $3\,\sigma$ & $95\%$ & $3\,\sigma$ \\
$\mathrm{Br}(t \to Zq)$ $(\gamma_\mu)$ &
  $1.3 \times 10^{-3}$ & $1.6 \times 10^{-3}$ &
  $1.1 \times 10^{-3}$ & $1.4 \times 10^{-3}$ &
  $8.3 \times 10^{-4}$ & $8.0 \times 10^{-4}$ \\
$\mathrm{Br}(t \to Zq)$ $(\sigma_{\mu \nu})$ &
  $3.9 \times 10^{-5}$ & $4.7 \times 10^{-5}$ &
  $3.2 \times 10^{-5}$ & $4.2 \times 10^{-5}$ &
  $2.5 \times 10^{-5}$ & $2.4 \times 10^{-5}$ \\
$\mathrm{Br}(t \to \gamma q)$ &
  $2.3 \times 10^{-5}$ & $2.8 \times 10^{-5}$ &
  $1.7 \times 10^{-5}$ & $2.2 \times 10^{-5}$ &
  $1.3 \times 10^{-5}$ & $1.2 \times 10^{-5}$
\end{tabular}
\caption{Limits on top FCN decays obtained from single top production at 800 GeV
with a reference luminosity of 100 fb$^{-1}$ for the three polarization options.
\label{tab:tc7}}
\end{small}
\end{center}
\end{table}

\begin{table}[htb]
\begin{center}
\begin{small}
\begin{tabular}{lcccccc}
& \multicolumn{2}{c}{No pol.} & \multicolumn{2}{c}{Pol. $e^-$}
& \multicolumn{2}{c}{Pol. $e^-$ $e^+$} \\
& $95\%$ & $3\,\sigma$ & $95\%$ & $3\,\sigma$ & $95\%$ & $3\,\sigma$ \\
$\mathrm{Br}(t \to Zq)$ $(\gamma_\mu)$ &
  $4.4 \times 10^{-4}$ & $5.9 \times 10^{-4}$ &
  $2.9 \times 10^{-4}$ & $4.3 \times 10^{-4}$ &
  $2.4 \times 10^{-4}$ & $2.3 \times 10^{-4}$ \\
$\mathrm{Br}(t \to Zq)$ $(\sigma_{\mu \nu})$ &
  $1.3 \times 10^{-5}$ & $1.7 \times 10^{-5}$ &
  $8.6 \times 10^{-6}$ & $1.3 \times 10^{-5}$ &
  $6.2 \times 10^{-6}$ & $7.0 \times 10^{-6}$ \\
$\mathrm{Br}(t \to \gamma q)$ &
  $7.8 \times 10^{-6}$ & $1.0 \times 10^{-5}$ &
  $4.5 \times 10^{-6}$ & $6.7 \times 10^{-6}$ &
  $3.7 \times 10^{-6}$ & $3.6 \times 10^{-6}$
\end{tabular}
\caption{Limits on top FCN decays obtained from single top production at 800
GeV with a luminosity of 500 fb$^{-1}$ for the three polarization options.
\label{tab:tc8}}
\end{small}
\end{center}
\end{table}

For integrated luminosities of 100 fb$^{-1}$ the limits on branching ratios of
decays mediated by non-renormalizable couplings are 1.5 times better than at 500
GeV, whereas the other limits are worse. Comparing the limits for 100 fb$^{-1}$
and 500 fb$^{-1}$ we notice that in the cases with polarization the limits for
500 fb$^{-1}$ are a factor of $3.3-3.8$ smaller instead of $\sqrt 5 \simeq 2.2$.
This improvement beyond the $1/\sqrt L$ scaling is a consequence of the small
backgrounds in these cases.

\section{Process $e^+ e^- \to t \bar q \gamma$}
\label{sec:4}
We begin the analysis of top decay processes with the more interesting case of
the $t \bar q \gamma$ signal. We study $t \bar t$ production with
$\bar t \to \gamma \bar q$ and $t$ decay to $W^+b$, that gives a final state
$l^+ \nu bj \gamma$, and sum the charge conjugate process. For equal values of
$\lambda_{tq}$, the process has a smaller cross-section than single top
production. The main reasons are: ({\em i\/}) top decays are insensitive to the
momentum factor $q^\nu$ of the $\sigma_{\mu \nu}$ coupling, ({\em ii\/})
the phase space for
the production of a top-antitop pair is smaller. Nevertheless, this process can
be useful to determine the nature of a FCN coupling involving the top quark,
since the final state signals a $\gamma tq$ coupling. The background is
$W^+ jj\gamma $ production with $W^+ \to l^+ \nu$ and a jet misidentified as a
$b$. As for single top production, we give our results for $\lambda_{tq}^v=1$,
$\lambda_{tq}^a=0$ but check that for other values of these parameters the
differences are of order $0.1\%$ and smaller that the Monte Carlo uncertainty
for the three polarization options, before and after kinematical cuts.

\begin{figure}[htb]
\begin{center}
\mbox{\epsfig{file=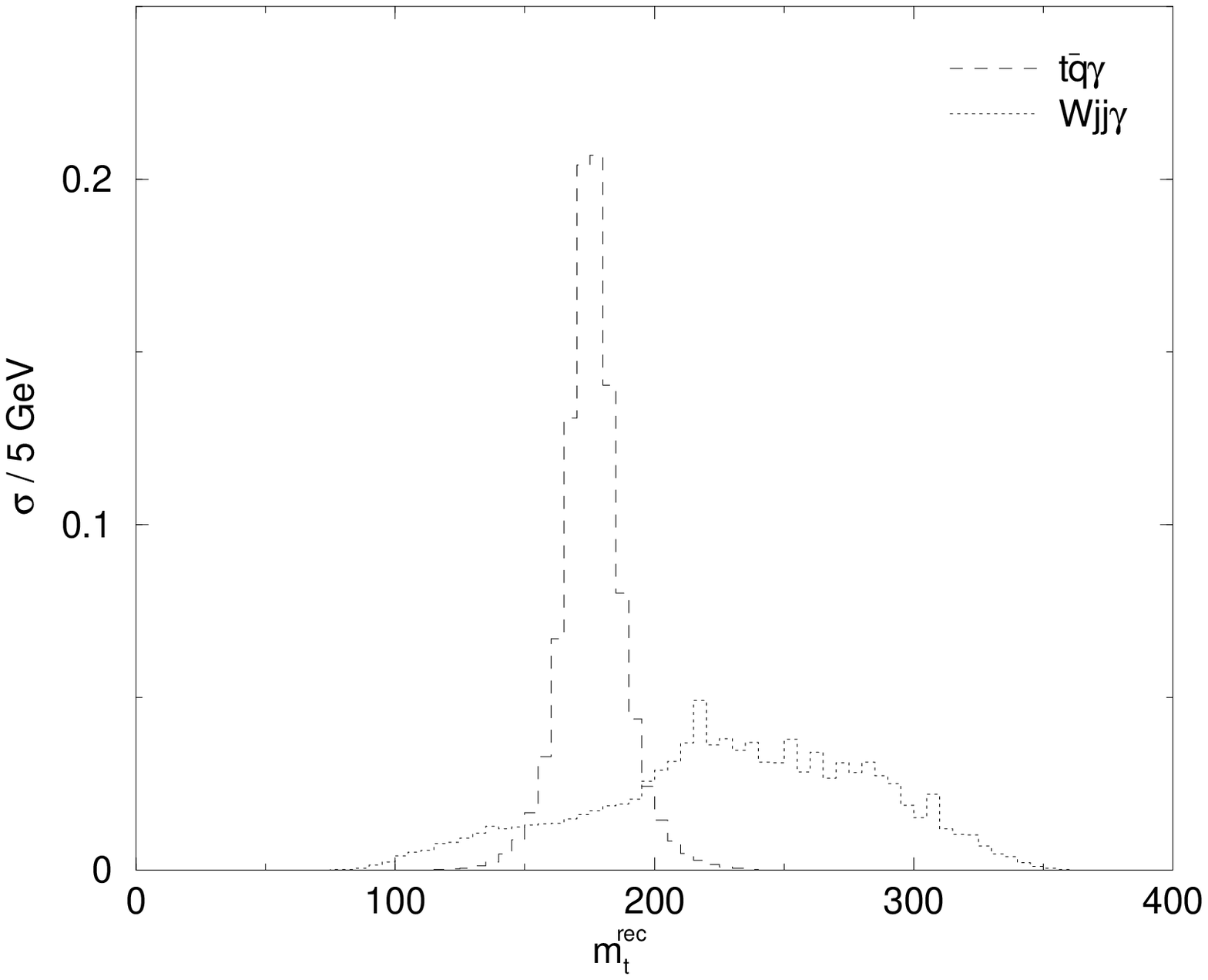,width=9cm}}
\end{center}
\caption{Reconstructed top mass $m_t^\mathrm{rec}$ distribution before
kinematical cuts for the $t \bar q \gamma$ signal and $W^+ jj\gamma$ background
at a CM energy of 500 GeV, without beam polarization. The cross-sections are
normalized to unity.
\label{fig:tca-mt}}
\end{figure}

\begin{figure}[htb]
\begin{center}
\mbox{\epsfig{file=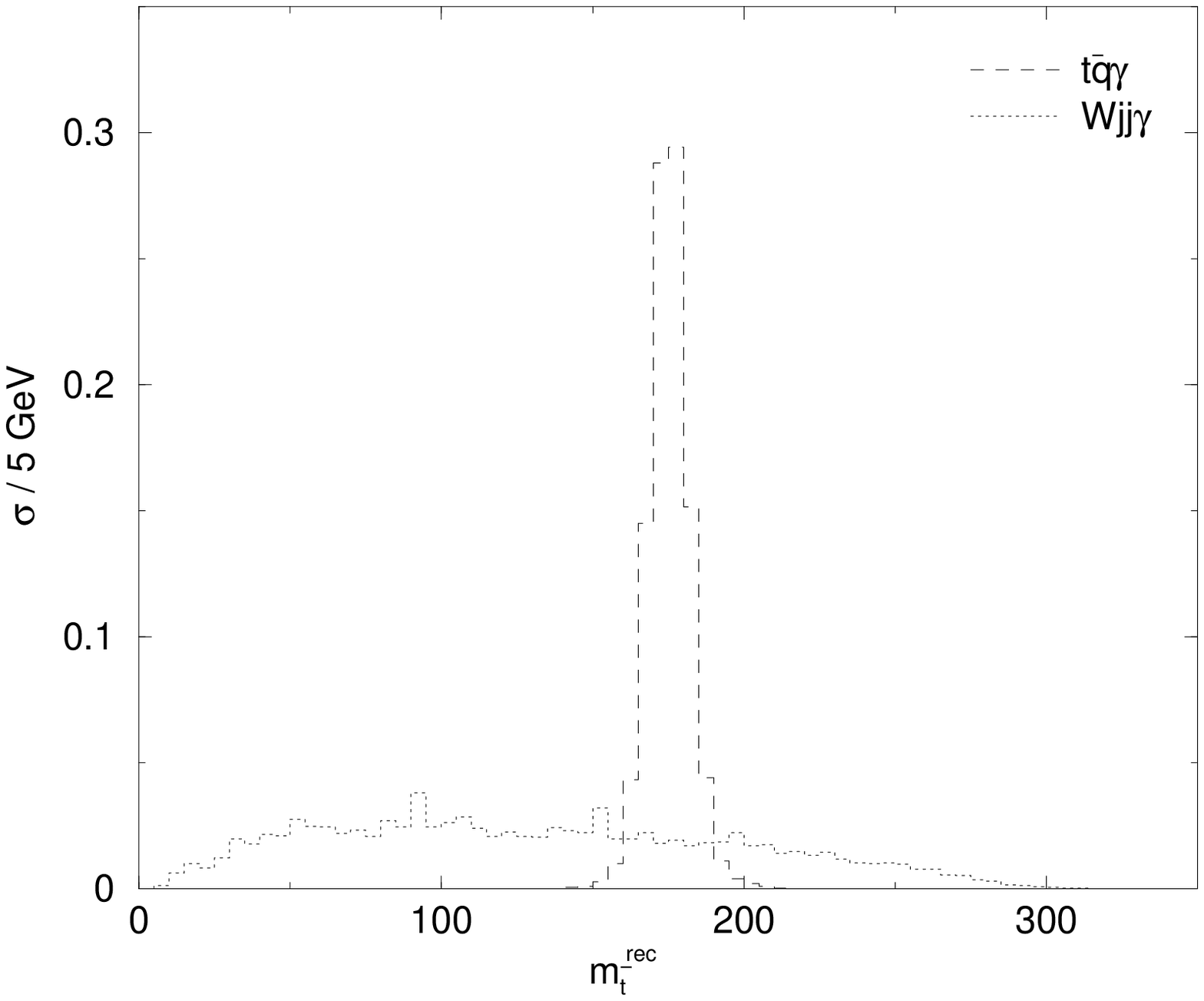,width=9cm}}
\end{center}
\caption{Reconstructed antitop mass $m_{\bar t}^\mathrm{rec}$ distribution
before kinematical cuts for the $t \bar q \gamma$ signal and $W^+ jj\gamma$
background at a CM energy of 500 GeV, without beam polarization. The
cross-sections are normalized to unity.
\label{fig:tca-mtb}}
\end{figure}

\begin{figure}[htb]
\begin{center}
\mbox{\epsfig{file=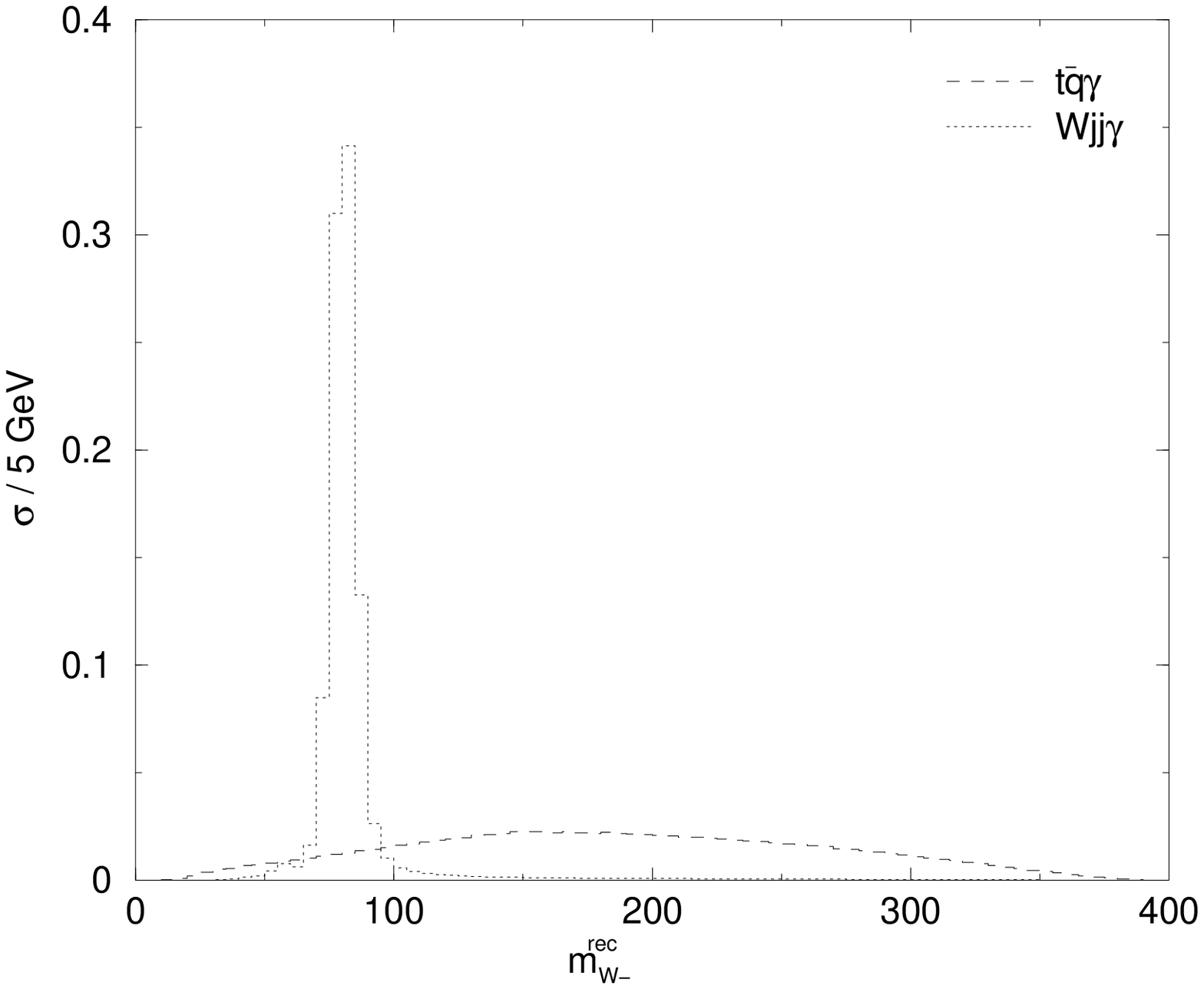,width=9cm}}
\end{center}
\caption{Reconstructed $W^-$ mass $M_{W^-}^\mathrm{rec}$ distribution before
kinematical cuts for the $t \bar q \gamma$ signal and $W^+ jj\gamma$ background
at a CM energy of 500 GeV, without beam polarization. The cross-sections are
normalized to unity.
\label{fig:tca-mw2}}
\end{figure}

The $t \bar q \gamma$ signal can be reconstructed in a similar way as $t \bar
q$. The $W^+$ momentum is the charged lepton momentum plus the missing momentum.
The invariant mass of the $W^+$ and one of the jets, $m_t^\mathrm{rec}$,
is consistent with the top mass, and the invariant mass of the photon and the
other jet, $m_{\bar t}^\mathrm{rec}$, is also consistent with $m_t$. Of the two
possible assignments, we choose the one minimizing
$(m_t^\mathrm{rec}-m_t)^2+(m_{\bar t}^\mathrm{rec}-m_t)^2$ and require a
$b$ tag on the jet that corresponds to the top quark. We plot the distributions
for both variables in Figs.~\ref{fig:tca-mt} and \ref{fig:tca-mtb}, for a CM
energy of 500 GeV without polarization. In the polarized case the distributions
are indistinguishable. We note that the $m_{\bar t}^\mathrm{rec}$ distribution
is more concentrated around $m_t$ because the energy resolution effects are less
important. The different behaviour of the $m_t^\mathrm{rec}$ distribution
of the background around $m_t$ is related to the cut $E_b > 45$. It is very
useful to consider also the invariant mass of the two jets,
$M_{W^-}^\mathrm{rec}$, which is $M_W$ for the background (see
Fig.~\ref{fig:tca-mw2}). We discuss the results for 500 GeV and 800 GeV
independently.

\subsection{Limits at 500 GeV}

We write the kinematical cuts for 500 GeV in Table~\ref{tab:tca1}. The cut on
$m_{\bar t}^\mathrm{rec}$ is more strict than the one on
$m_t^\mathrm{rec}$ because the reconstruction of the antitop mass is
better. The cross-sections before and after cuts are gathered in
Table~\ref{tab:tca2}. Note that we normalize the signal to $\lambda_{tq} = 0.04$
instead of $\lambda_{tq} = 0.02$ as in the previous Section because the
cross-sections are much smaller.

\begin{table}[htb]
\begin{center}
\begin{tabular}{ccc}
Variable & 500 GeV cut \\
$m_t^\mathrm{rec}$ & 150--200 \\
$m_{\bar t}^\mathrm{rec}$ & 160--190 \\
$M_{W^-}^\mathrm{rec}$ & $<65$ or $>95$
\end{tabular}
\caption{Kinematical cuts for the $t \bar q \gamma$ signal and the three
polarization options at a CM energy of 500 GeV. The masses are in GeV.
\label{tab:tca1}}
\end{center}
\end{table}

\begin{table}[htb]
\begin{center}
\begin{tabular}{lcccccc}
& \multicolumn{2}{c}{No pol.} & \multicolumn{2}{c}{Pol. $e^-$}
& \multicolumn{2}{c}{Pol. $e^-$ $e^+$} \\
& before & after & before & after & before & after \\[-0.2cm]
& cuts & cuts & cuts & cuts & cuts & cuts \\
$t \bar q \gamma +\bar tq \gamma$
                 & 0.0745 & 0.0631 & 0.0515 & 0.0429 & 0.0653 & 0.0543 \\
$W^\pm jj\gamma$ & 0.639  & 0.0014 & 0.144  & $3.1 \times 10^{-4}$ & 0.0956 &
 $2.0 \times 10^{-4}$
\end{tabular}
\caption{Cross-sections (in fb) before and after the kinematical cuts in
Table~\ref{tab:tca1} for the  $t \bar q \gamma$ signal and background at a CM
energy of 500 GeV, for the three polarization options. We include $b$ tagging
efficiencies and use $\lambda_{tq} = 0.04$.
\label{tab:tca2}}
\end{center}
\end{table}

The use of polarization is not as useful as for single top production. Although
it reduces the $W^+ jj\gamma$ cross-section up to a factor of 6, this background
is already tiny without polarization, and there is little advantage in reducing
it further. Moreover, the signal cross-section also decreases, and the limits
obtained are in some cases worse (see Tables~\ref{tab:tca3} and \ref{tab:tca4}).
The limits from the $t \bar q \gamma$ signal are in all cases worse than those
obtained from single top production. Polarization of
$e^-$ only gives worse results, but $e^-$ and $e^+$ polarization improves the
$3\,\sigma$ discovery limits.

\begin{table}[htb]
\begin{center}
\begin{small}
\begin{tabular}{lcccccc}
& \multicolumn{2}{c}{No pol.} & \multicolumn{2}{c}{Pol. $e^-$}
& \multicolumn{2}{c}{Pol. $e^-$ $e^+$} \\
& $95\%$ & $3\,\sigma$ & $95\%$ & $3\,\sigma$ & $95\%$ & $3\,\sigma$ \\
$\mathrm{Br}(t \to \gamma q)$ &
  $3.3 \times 10^{-4}$ & $3.2 \times 10^{-4}$ &
  $5.0 \times 10^{-4}$ & $3.2 \times 10^{-4}$ &
  $4.0 \times 10^{-4}$ & $2.6 \times 10^{-4}$
\end{tabular}
\end{small}
\caption{Limits on top FCN decays obtained from the $t \bar q \gamma$ process at
500 GeV with a reference luminosity of 100 fb$^{-1}$ for the three polarization
options.
\label{tab:tca3}}
\end{center}
\end{table}

\begin{table}[htb]
\begin{center}
\begin{small}
\begin{tabular}{lcccccc}
& \multicolumn{2}{c}{No pol.} & \multicolumn{2}{c}{Pol. $e^-$}
& \multicolumn{2}{c}{Pol. $e^-$ $e^+$} \\
& $95\%$ & $3\,\sigma$ & $95\%$ & $3\,\sigma$ & $95\%$ & $3\,\sigma$ \\
$\mathrm{Br}(t \to \gamma q)$ &
  $9.9 \times 10^{-5}$ & $1.3 \times 10^{-4}$ &
  $1.6 \times 10^{-4}$ & $1.6 \times 10^{-4}$ &
  $1.3 \times 10^{-4}$ & $8.3 \times 10^{-5}$
\end{tabular}
\end{small}
\caption{Limits on top FCN decays obtained from the $t \bar q \gamma$ process at
500 GeV with a luminosity of 300 fb$^{-1}$ for the three polarization options.
\label{tab:tca4}}
\end{center}
\end{table}

\subsection{Limits at 800 GeV}

For 800 GeV we perform the loose cuts for the top, antitop and $W^-$
reconstruction in Table~\ref{tab:tca5}, because the background is smaller than
at 500 GeV. The signal cross-section also decreases in spite of the $q^\nu$
factor in the $\sigma_{\mu \nu}$ coupling, and the limits obtained are worse.
The cross-sections before and after cuts can be read in Table~\ref{tab:tca6},
and the limits obtained for 100 fb$^{-1}$ and 500 fb$^{-1}$ in
Tables~\ref{tab:tca7} and \ref{tab:tca8} respectively. The same comments as for
500 GeV apply in this case. We can see that the limits for 800 GeV and 500
fb$^{-1}$ are similar but worse than those obtained for 500 GeV and 300
fb$^{-1}$.

\begin{table}[htb]
\begin{center}
\begin{tabular}{ccc}
Variable & 800 GeV cut \\
$m_t^\mathrm{rec}$ & 130--220 \\
$m_{\bar t}^\mathrm{rec}$ & 150--200 \\
$M_{W^-}^\mathrm{rec}$ & $<60$ or $>100$
\end{tabular}
\caption{Kinematical cuts for the $t \bar q \gamma$ signal and the three
polarization options at a CM energy of 800 GeV. The masses are in GeV.
\label{tab:tca5}}
\end{center}
\end{table}

\begin{table}[htb]
\begin{center}
\begin{tabular}{lcccccc}
& \multicolumn{2}{c}{No pol.} & \multicolumn{2}{c}{Pol. $e^-$}
& \multicolumn{2}{c}{Pol. $e^-$ $e^+$} \\
& before & after & before & after & before & after \\[-0.2cm]
& cuts & cuts & cuts & cuts & cuts & cuts \\
$t \bar q \gamma +\bar tq \gamma$
                 & 0.0350 & 0.0327 & 0.0246 & 0.0227 & 0.0314 & 0.0289 \\
$W^\pm jj\gamma$ & 0.437  & $8.2 \times 10^{-4}$ & 0.0959 &
 $1.8 \times 10^{-4}$ & 0.0613 & $1.1 \times 10^{-4}$
\end{tabular}
\caption{Cross-sections (in fb) before and after the kinematical cuts in
Table~\ref{tab:tca5} for the $t \bar q \gamma$ signal and background at a CM
energy of 800 GeV, for the three polarization options. We include $b$ tagging
efficiencies and use $\lambda_{tq} = 0.04$.
\label{tab:tca6}}
\end{center}
\end{table}

\begin{table}[htb]
\begin{center}
\begin{small}
\begin{tabular}{lcccccc}
& \multicolumn{2}{c}{No pol.} & \multicolumn{2}{c}{Pol. $e^-$}
& \multicolumn{2}{c}{Pol. $e^-$ $e^+$} \\
& $95\%$ & $3\,\sigma$ & $95\%$ & $3\,\sigma$ & $95\%$ & $3\,\sigma$ \\
$\mathrm{Br}(t \to \gamma q)$ &
  $6.5 \times 10^{-4}$ & $6.3 \times 10^{-4}$ &
  $9.5 \times 10^{-4}$ & $6.1 \times 10^{-4}$ &
  $7.5 \times 10^{-4}$ & $4.8 \times 10^{-4}$
\end{tabular}
\end{small}
\caption{Limits on top FCN decays obtained from the $t \bar q \gamma$ process
at 800 GeV with a reference luminosity of 100 fb$^{-1}$ for the three
polarization options.
\label{tab:tca7}}
\end{center}
\end{table}

\begin{table}[htb]
\begin{center}
\begin{small}
\begin{tabular}{lcccccc}
& \multicolumn{2}{c}{No pol.} & \multicolumn{2}{c}{Pol. $e^-$}
& \multicolumn{2}{c}{Pol. $e^-$ $e^+$} \\
& $95\%$ & $3\,\sigma$ & $95\%$ & $3\,\sigma$ & $95\%$ & $3\,\sigma$ \\
$\mathrm{Br}(t \to \gamma q)$ &
  $1.2 \times 10^{-4}$ & $1.5 \times 10^{-4}$ &
  $1.9 \times 10^{-4}$ & $1.8 \times 10^{-4}$ &
  $1.5 \times 10^{-4}$ & $9.4 \times 10^{-5}$
\end{tabular}
\end{small}
\caption{Limits on top FCN decays obtained from the $t \bar q \gamma$ process
at 800 GeV with a luminosity of 500 fb$^{-1}$ for the three polarization
options.
\label{tab:tca8}}
\end{center}
\end{table}

\section{Process $e^+ e^- \to t \bar qZ$}
\label{sec:5}
In this Section we study $t \bar t$ production with $\bar t \to Z \bar q$ and
$t \to W^+ b$, that gives a final state $l^+ \nu b \bar q l'^+ l'^-$. This
signal is analogous to $t \bar q \gamma$, but with the disadvantage that the
partial width $\mathrm{Br}(Z \to l'^+ l'^-) = 0.067$ considerably decreases the
cross-sections for the signal and background. Comparing with single top
production, we find that for equal values of $X_{tq}$, $\kappa_{tq}$ the
cross-section for $t \bar q Z$ is much smaller, mainly for the inclusion of the
$Z$ partial width, and for the smaller phase space also. In addition,
for $\sigma_{\mu \nu}$ couplings the $t \bar qZ$ signal does not have an
enhancement from the $q^\nu$ factor in the vertex. We give our results for
$\gamma_\mu$ and $\sigma_{\mu \nu}$ couplings using for definiteness
$x_{tq}^L = x_{tq}^R$ and $\kappa_{tq}^v=1$, $\kappa_{tq}^a=0$, respectively.
We check that the differences with other chiralities are insignificant before
and after kinematical cuts for the three polarization options.

\begin{figure}[htb]
\begin{center}
\mbox{\epsfig{file=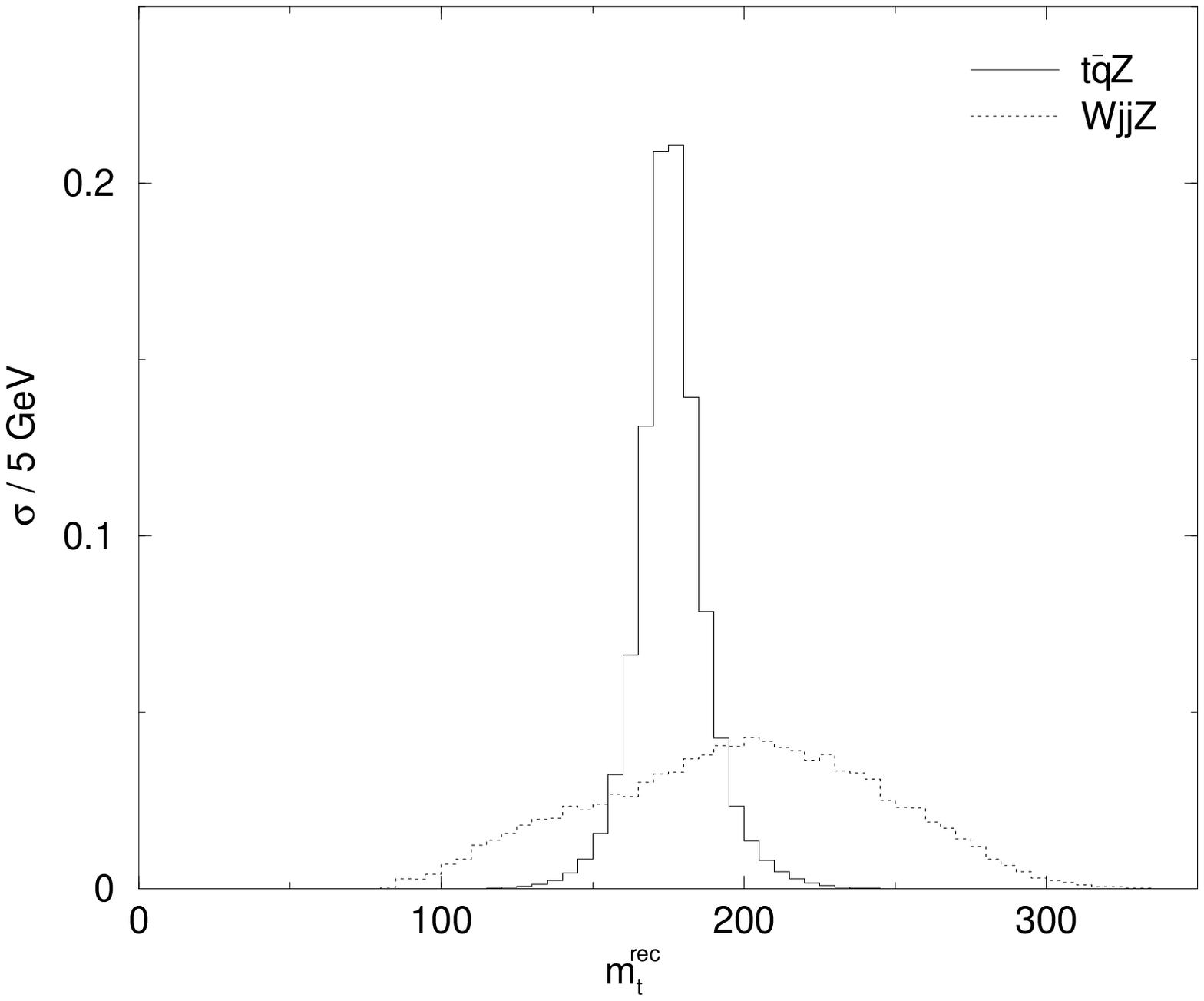,width=9cm}}
\end{center}
\caption{Reconstructed top mass $m_t^\mathrm{rec}$ distribution before
kinematical cuts for the $t \bar q Z$ signal and $W^+ jjZ$ background at a CM
energy of 500 GeV, without beam polarization. The cross-sections are normalized
to unity.
\label{fig:tcz-mt}}
\end{figure}

\begin{figure}[htb]
\begin{center}
\mbox{\epsfig{file=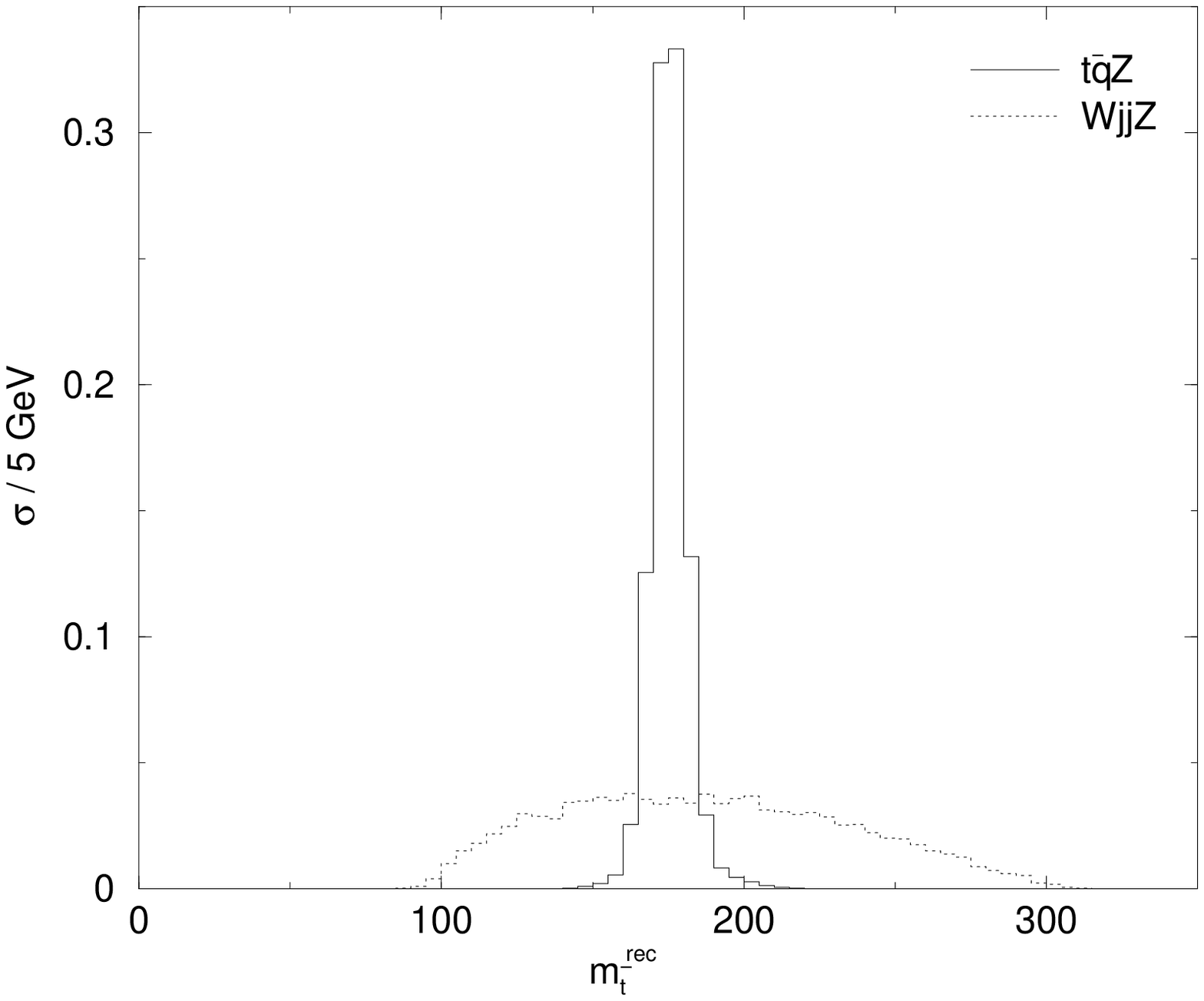,width=9cm}}
\end{center}
\caption{Reconstructed antitop mass $m_{\bar t}^\mathrm{rec}$ distribution
before kinematical cuts for the $t \bar qZ$ signal and $W^+ jjZ$ background at a
CM energy of 500 GeV, without beam polarization. The cross-sections are
normalized to unity.
\label{fig:tcz-mtb}}
\end{figure}

\begin{figure}[htb]
\begin{center}
\mbox{\epsfig{file=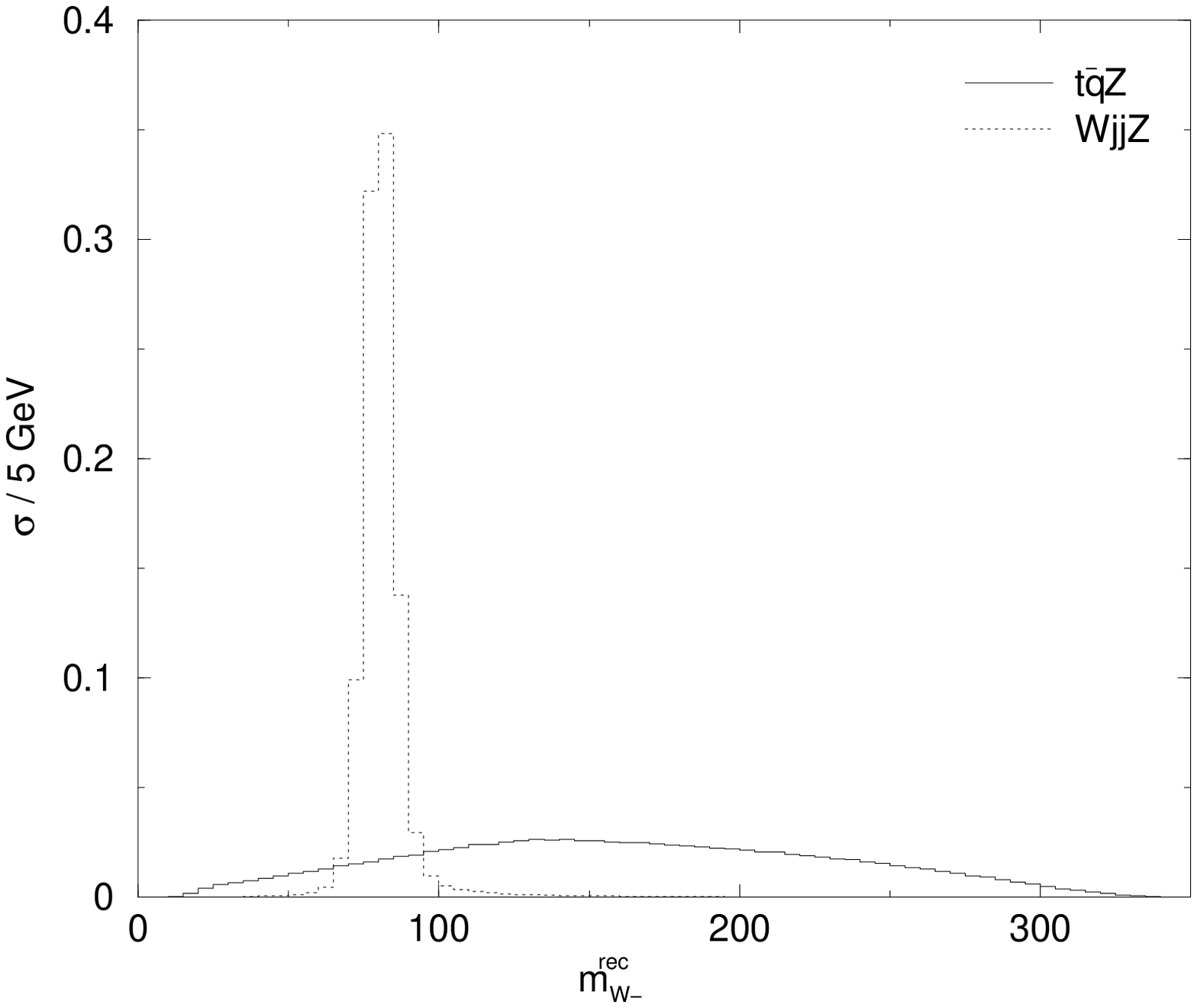,width=9cm}}
\end{center}
\caption{Reconstructed $W^-$ mass $M_{W^-}^\mathrm{rec}$ distribution before
kinematical cuts for the $t \bar qZ$ signal and $W^+ jjZ$ background at a CM
energy of 500 GeV, without beam polarization. The cross-sections are normalized
to unity.
\label{fig:tcz-mw2}}
\end{figure}

The background is $W^+ jjZ$ production with $W^+ \to l^+ \nu$, $Z \to l'^+ l'^-$
and a $b$ mistag. We reconstruct the signal and background as follows. Of the
two positively charged leptons, one results from the $W^+$ decay and it has with
the neutrino (reconstructed from the missing momentum)  an invariant mass
$M_{W^+}^\mathrm{rec}$ consistent with $M_W$. The other one and the negative
charge lepton have an invariant mass $M_Z^\mathrm{rec}$ close to $M_Z$. If the
two positive leptons have different flavours the assignment is straightforward,
but if not we choose the pairing that minimizes
$(M_{W^+}^\mathrm{rec}-M_W)^2+(M_Z^\mathrm{rec}-M_Z)^2$. Then, we
reconstruct the top and antitop masses as for the $t \bar q \gamma$ signal
replacing the photon momentum with the $Z$ momentum. The $W^-$ reconstruction
for the background is also similar. These distributions are plotted in
Figs.~\ref{fig:tcz-mt}--\ref{fig:tcz-mw2}.

\subsection{Limits at 500 GeV}

Since the background is so small for this signal, we only apply very loose
kinematical cuts for the top, antitop and $W^-$ reconstruction. These can be
read in Table~\ref{tab:tcz1}, and the cross-sections before and after these cuts
in Table~\ref{tab:tcz2}. Note that we use a different normalization, $X_{tq} =
0.2$, $\kappa_{tq} = 0.2$, because the cross-sections are very small.

\begin{table}[htb]
\begin{center}
\begin{tabular}{ccc}
Variable & 500/800 GeV cut \\
$m_t^\mathrm{rec}$ & 130--220 \\
$m_{\bar t}^\mathrm{rec}$ & 150--200 \\
$M_{W^-}^\mathrm{rec}$ & $<70$ or $>90$
\end{tabular}
\caption{Kinematical cuts for the $t \bar qZ$ signal and the three polarization
options at CM energies of 500 and 800 GeV. The masses are in GeV.
\label{tab:tcz1}}
\end{center}
\end{table}

\begin{table}[htb]
\begin{center}
\begin{tabular}{lcccccc}
& \multicolumn{2}{c}{No pol.} & \multicolumn{2}{c}{Pol. $e^-$}
& \multicolumn{2}{c}{Pol. $e^-$ $e^+$} \\
& before & after & before & after & before & after \\[-0.2cm]
& cuts & cuts & cuts & cuts & cuts & cuts \\
$t \bar qZ +\bar tqZ$ $(\gamma_\mu)$
            & 0.114  & 0.105  & 0.0784 & 0.0720 & 0.0995 & 0.0912 \\
$t \bar qZ +\bar tqZ$ $(\sigma_{\mu \nu})$
            & 0.0877 & 0.0809 & 0.0604 & 0.0555 & 0.0766 & 0.0703 \\
$W^\pm jjZ$ & 0.0059 & $1.0 \times 10^{-4}$ & 0.0013 & $2.4 \times 10^{-5}$ &
$8.9 \times 10^{-4}$ & $1.6 \times 10^{-5}$
\end{tabular}
\caption{Cross-sections (in fb) before and after the kinematical cuts in
Table~\ref{tab:tcz1} for the $t \bar qZ$ signal and background at a CM energy of
500 GeV, for the three polarization options. We include $b$ tagging efficiencies
and use $X_{tq} = 0.2$, $\kappa_{tq} = 0.2$.
\label{tab:tcz2}}
\end{center}
\end{table}

Polarization is not as useful as for $t\bar q$ production, and the behaviour is
similar as for the $t \bar q \gamma$ signal.  This is reflected in the limits in
Tables~\ref{tab:tcz3} and \ref{tab:tcz4}, where we find that polarization in
some cases gives worse results. Note that $3\,\sigma$ discovery limits do not
follow the same pattern for 100 fb$^{-1}$ and 300 fb$^{-1}$ due to the
discreteness of Poisson statistics.
These limits are much worse than those obtained from $t\bar q$ production, one
order of magnitude worse for $\gamma_\mu$ couplings and two orders for
$\sigma_{\mu \nu}$ couplings. In fact, this process would only be useful if a
FCN top decay is detected with $\mathrm{Br}(t \to Zq) \sim 10^{-3}$. In such
case, it would help to determine the nature of the top anomalous coupling.
Besides, it is interesting to notice that the limits for $\gamma_\mu$ and
$\sigma_{\mu \nu}$ couplings are remarkably similar. This confirms that this
process is not sensitive to the $q^\nu$ factor of the $\sigma_{\mu \nu}$ vertex.

\begin{table}[htb]
\begin{center}
\begin{small}
\begin{tabular}{lcccccc}
& \multicolumn{2}{c}{No pol.} & \multicolumn{2}{c}{Pol. $e^-$}
& \multicolumn{2}{c}{Pol. $e^-$ $e^+$} \\
& $95\%$ & $3\,\sigma$ & $95\%$ & $3\,\sigma$ & $95\%$ & $3\,\sigma$ \\
$\mathrm{Br}(t \to Zq)$ $(\gamma_\mu)$ &
  $5.4 \times 10^{-3}$ & $3.5 \times 10^{-3}$ &
  $8.0 \times 10^{-3}$ & $2.6 \times 10^{-3}$ &
  $6.3 \times 10^{-3}$ & $2.0 \times 10^{-3}$ \\
$\mathrm{Br}(t \to Zq)$ $(\sigma_{\mu \nu})$ &
  $5.7 \times 10^{-3}$ & $3.7 \times 10^{-3}$ &
  $8.3 \times 10^{-3}$ & $2.7 \times 10^{-3}$ &
  $6.5 \times 10^{-3}$ & $2.1 \times 10^{-3}$
\end{tabular}
\end{small}
\caption{Limits on top FCN decays obtained from the $t \bar qZ$ process at 500
GeV with a reference luminosity of 100 fb$^{-1}$ for the three polarization
options.
\label{tab:tcz3}}
\end{center}
\end{table}

\begin{table}[htb]
\begin{center}
\begin{small}
\begin{tabular}{lcccccc}
& \multicolumn{2}{c}{No pol.} & \multicolumn{2}{c}{Pol. $e^-$}
& \multicolumn{2}{c}{Pol. $e^-$ $e^+$} \\
& $95\%$ & $3\,\sigma$ & $95\%$ & $3\,\sigma$ & $95\%$ & $3\,\sigma$ \\
$\mathrm{Br}(t \to Zq)$ $(\gamma_\mu)$ &
  $1.8 \times 10^{-3}$ & $1.2 \times 10^{-3}$ &
  $2.7 \times 10^{-3}$ & $1.7 \times 10^{-3}$ &
  $2.1 \times 10^{-3}$ & $1.4 \times 10^{-3}$ \\
$\mathrm{Br}(t \to Zq)$ $(\sigma_{\mu \nu})$ &
  $1.9 \times 10^{-3}$ & $1.2 \times 10^{-3}$ &
  $2.8 \times 10^{-3}$ & $1.8 \times 10^{-3}$ &
  $2.2 \times 10^{-3}$ & $1.4 \times 10^{-3}$
\end{tabular}
\end{small}
\caption{Limits on top FCN decays obtained from the $t \bar qZ$ process at 500
GeV with a luminosity of 300 fb$^{-1}$ for the three polarization options.
\label{tab:tcz4}}
\end{center}
\end{table}

\subsection{Limits at 800 GeV}

For 800 GeV we use the same set of cuts in Table~\ref{tab:tcz1} for the top,
antitop and $W^-$ reconstruction, and obtain the cross-sections in
Table~\ref{tab:tcz6}. We notice that the background before cuts is larger than
at 500 GeV (remember that it is dominated by $W^+ W^- Z$ production, and its
cross-section 
increases until CM energies around 900 GeV), but after the veto cut to remove
events with on-shell $W^-$ it becomes smaller as expected. We collect the limits
obtained in Tables~\ref{tab:tcz7} and \ref{tab:tcz8}. The same comments as for
the 500 GeV analysis apply.

\begin{table}[htb]
\begin{center}
\begin{tabular}{lcccccc}
& \multicolumn{2}{c}{No pol.} & \multicolumn{2}{c}{Pol. $e^-$}
& \multicolumn{2}{c}{Pol. $e^-$ $e^+$} \\
& before & after & before & after & before & after \\[-0.2cm]
& cuts & cuts & cuts & cuts & cuts & cuts \\
$t \bar qZ +\bar tqZ$ $(\gamma_\mu)$
            & 0.0523 & 0.0496 & 0.0367 & 0.0345 & 0.0467 & 0.0439 \\
$t \bar qZ +\bar tqZ$ $(\sigma_{\mu \nu})$
            & 0.0387 & 0.0367 & 0.0272 & 0.0255 & 0.0346 & 0.0325  \\
$W^\pm jjZ$ & 0.0091 & $2.3 \times 10^{-5}$ & 0.0020 & $5.0 \times 10^{-6}$ &
0.0012 & $3.3 \times 10^{-6}$
\end{tabular}
\caption{Cross-sections (in fb) before and after the kinematical cuts in
Table~\ref{tab:tcz1} for the $t \bar qZ$ signal and background at a CM energy of
800 GeV, for the three polarization options. We include $b$ tagging efficiencies
and use $X_{tq} = 0.04$, $\kappa_{tq} = 0.04$.
\label{tab:tcz6}}
\end{center}
\end{table}

\begin{table}[htb]
\begin{center}
\begin{small}
\begin{tabular}{lcccccc}
& \multicolumn{2}{c}{No pol.} & \multicolumn{2}{c}{Pol. $e^-$}
& \multicolumn{2}{c}{Pol. $e^-$ $e^+$} \\
& $95\%$ & $3\,\sigma$ & $95\%$ & $3\,\sigma$ & $95\%$ & $3\,\sigma$ \\
$\mathrm{Br}(t \to Zq)$ $(\gamma_\mu)$ &
  $1.2 \times 10^{-2}$ & $3.7 \times 10^{-3}$ &
  $1.7 \times 10^{-2}$ & $5.4 \times 10^{-3}$ &
  $1.3 \times 10^{-2}$ & $4.2 \times 10^{-3}$ \\
$\mathrm{Br}(t \to Zq)$ $(\sigma_{\mu \nu})$ &
  $1.3 \times 10^{-2}$ & $4.0 \times 10^{-3}$ &
  $1.8 \times 10^{-2}$ & $5.8 \times 10^{-3}$ &
  $1.4 \times 10^{-2}$ & $4.6 \times 10^{-3}$
\end{tabular}
\end{small}
\caption{Limits on top FCN decays obtained from the $t \bar qZ$ process at 800
GeV with a reference luminosity of 100 fb$^{-1}$ for the three polarization
options.
\label{tab:tcz7}}
\end{center}
\end{table}

\begin{table}[htb]
\begin{center}
\begin{small}
\begin{tabular}{lcccccc}
& \multicolumn{2}{c}{No pol.} & \multicolumn{2}{c}{Pol. $e^-$}
& \multicolumn{2}{c}{Pol. $e^-$ $e^+$} \\
& $95\%$ & $3\,\sigma$ & $95\%$ & $3\,\sigma$ & $95\%$ & $3\,\sigma$ \\
$\mathrm{Br}(t \to Zq)$ $(\gamma_\mu)$ &
  $2.3 \times 10^{-3}$ & $1.5 \times 10^{-3}$ &
  $3.3 \times 10^{-3}$ & $1.1 \times 10^{-3}$ &
  $2.6 \times 10^{-3}$ & $8.4 \times 10^{-4}$ \\
$\mathrm{Br}(t \to Zq)$ $(\sigma_{\mu \nu})$ &
  $2.5 \times 10^{-3}$ & $1.6 \times 10^{-3}$ &
  $3.6 \times 10^{-3}$ & $1.2 \times 10^{-3}$ &
  $2.8 \times 10^{-3}$ & $9.1 \times 10^{-4}$
\end{tabular}
\end{small}
\caption{Limits on top FCN decays obtained from the $t \bar qZ$ process at 800
GeV with a luminosity of 500 fb$^{-1}$ for the three polarization options.
\label{tab:tcz8}}
\end{center}
\end{table}

\section{Summary}
\label{sec:6}
We have studied the most important signals of top FCN couplings to the $Z$ boson
and the photon that can be observed at a future $e^+ e^-$ collider like TESLA.
These are single top production $e^+ e^- \to t \bar q$, and rare top decays
$e^+ e^- \to t\bar t \to t\bar q \gamma$, $e^+ e^- \to t\bar t \to t\bar qZ$.
We have discussed three beam polarization options: no polarization, $80\%$ $e^-$
polarization and $80\%$ $e^-$, $45\%$ $e^+$ polarization, for the two planned
energies of 500 GeV and 800 GeV. In the following we summarize the differences
among the signals and the influence of polarization and CM energy.

{\em Single top production versus top decays}.
Top decay signals are cleaner than single top production. This can be
understood since the top decay signals $W^+ bjV$ ($V=Z,\gamma$) have the
enhancement over their background $W^+ jjV$ of two on-shell particles, the top
and the antitop, whereas single top production has only the enhancement due to
the top on-shell and the $\sigma_{\mu \nu}$ coupling if that is the case. For
instance, we can compare the values after kinematical cuts for 500 GeV without
polarization for $\gamma tq$ couplings. We see that the $S/B$ ratio for the
$t \bar q \gamma$ signal (after rescaling to $\lambda_{tq} = 0.02$ as was 
assumed for single top production) is equal to12, and for $t \bar q$ it 
equals 4.

On the other hand, the cross-section for single top production is larger than
for top decays for equal values of the parameters. For our previous example,
$\sigma(t \bar q + \bar t q) = 0.288$ fb,
$\sigma(t \bar q \gamma + \bar t q \gamma) = 0.0158$ fb. The reasons are:
({\em i\/}) $t \bar q$ production is enhanced by the $q^\nu$ factor of the
$\sigma_{\mu \nu}$ vertex, whereas $t \bar q \gamma$ is not; ({\em ii\/})
Phase space for the
production of a $t \bar t$ pair is smaller than for $t \bar q$. This makes the
limits from $t \bar q$ production 4 times better for an integrated luminosity of
300 fb$^{-1}$.
However, if a positive signal of a $Vtq$ coupling is
discovered through single top production, top decays can help to determine the
nature of the coupling involved, {\em i.e.} whether it involves the photon or
the $Z$ boson, if $\mathrm{Br}(t \to Vq) \sim 10^{-4}$ or larger.

For $Ztq$ couplings similar comments apply. The top decay signals are cleaner,
especially for $\gamma_\mu$ couplings,
but the cross-sections are much smaller due to the leptonic partial width of the
$Z$, $\mathrm{Br}(Z \to l'^+ l'^-) = 0.067$. The limits obtained for
$\gamma_\mu$ couplings are one order of magnitude worse, and those for
$\sigma_{\mu \nu}$ couplings two orders of magnitude. Thus, the $t \bar qZ$
process is useful only if a signal with $\mathrm{Br}(t \to Vq) \sim 10^{-3}$ is
detected.

{\em Influence of beam polarization}.
Polarization is very useful to improve the limits from single top production.
In Table~\ref{tab:tc2} we can observe that for a CM energy of 500 GeV the use
of $80\%$ $e^-$ polarization decreases the background by a factor of 4.8 while
keeping $90\%$ of the signal, and additional $e^+$ polarization of $45\%$
decreases the background by a factor of 8.1 and increases the signal $17\%$
with respect to the values without polarization. The effect is similar at 800
GeV. $e^-$, $e^+$ polarization improves the $3\,\sigma$ discovery limits on
$\gamma_\mu$ couplings at 500 GeV with 300 fb$^{-1}$ by a factor of 3, and the
$3\,\sigma$ limits on $\sigma_{\mu \nu}$ couplings at 800 GeV with 500 fb$^{-1}$
by a factor of 2.6. The luminosities required to obtain the same results without
polarization would be 2100 fb$^{-1}$ and 3000 fb$^{-1}$, respectively.

For top decay signals polarization is not as useful, because the backgrounds are
already very small for unpolarized beams, and the luminosities required to
glimpse the potential improvement would exceed 1000 fb$^{-1}$. In addition,
the signal cross-sections decrease $10-20\%$, in contrast to single top
production. However, $e^-$ and $e^+$ polarization still gives an improvement in
the $\gamma tq$ coupling $3 \,\sigma$ discovery limits at 500 GeV with 300
fb$^{-1}$ of a factor of 1.6. This would be equivalent to double the luminosity
without polarization.

{\em Influence of centre of mass energy}.
The increase in CM energy from 500 GeV to 800 GeV enhances the sensitivity of
single top production to $\sigma_{\mu \nu}$ couplings. This is because the
signal cross-sections do not decrease (for the photon it even increases
slightly) whereas the background is less than one half at 800 GeV. An $e^+ e^-$
collider with 800 GeV and a reference luminosity of 100 fb$^{-1}$ is sensitive
to top rare decays mediated by these vertices with branching ratios $1.5-2$
times smaller than one with 500 GeV and the same luminosity. Of course, the
higher luminosity at 800 GeV has also to be taken into account, and then this
energy is best suited to perform searches for these vertices.

For normalizable $\gamma_\mu$ couplings the signal cross-sections decrease for
800 GeV as expected, and thus the sensitivity is worse, even after taking into
account the luminosity increase. More surprisingly, in top decays the limits are
worse for the three types of couplings, because top decays are not sensitive to
the $q^\nu$ factor of the $\sigma_{\mu \nu}$ vertex. Hence, to search for
$\gamma_\mu$ couplings in single top production and for all FCN coupling
searches in top decays the CM energy of 500 GeV is more adequate and gives the
best results.

{\em Conclusions}.
We compare the best limits on anomalous couplings that can be
obtained at TESLA and LHC. To obtain the values for LHC we rescale the data
from the literature to a $b$ tagging efficiency of $50\%$ and keep the average
mistagging rate used of $1\%$ for other jets, which is somewhat optimistic. The
best LHC limits on $Vtc$ couplings come from top decays, whereas the best ones
on $Vtu$ couplings are from single top production.  The LHC limit on
$\mathrm{Br}(t \to Zc)$ with $\sigma_{\mu \nu}$ couplings is estimated to be
similar to the one with $\gamma_\mu$ couplings having in mind the similar
result observed in
Section~\ref{sec:5}. We assume one year of running time in all the cases, that
is, 100 fb$^{-1}$ for LHC, 300 fb$^{-1}$ for TESLA at 500 GeV and 500 fb$^{-1}$
for TESLA at 800 GeV. We use the statistical estimators explained in
Section~\ref{sec:2}.

\begin{table}[htb]
\begin{center}
\begin{tabular}{ccccc}
& \multicolumn{2}{c}{LHC} & \multicolumn{2}{c}{TESLA} \\
& $95\%$ & $3\,\sigma$ & $95\%$ & $3\,\sigma$ \\
$\mathrm{Br}(t \to Zu)$ $(\gamma_\mu)$ &
  $6.2 \times 10^{-5}$ & $8.0 \times 10^{-5}$ &
  $1.9 \times 10^{-4}$ & $2.2 \times 10^{-4}$ \\
$\mathrm{Br}(t \to Zc)$ $(\gamma_\mu)$ &
  $7.1 \times 10^{-5}$ & $1.0 \times 10^{-4}$ &
  $1.9 \times 10^{-4}$ & $2.2 \times 10^{-4}$ \\
$\mathrm{Br}(t \to Zu)$ $(\sigma_{\mu \nu})$ &
  $1.8 \times 10^{-5}$ & $2.3 \times 10^{-5}$ &
  $6.2 \times 10^{-6}$ & $7.0 \times 10^{-6}$ \\
$\mathrm{Br}(t \to Zc)$ $(\sigma_{\mu \nu})$ &
  $7.1 \times 10^{-5}$ & $1.0 \times 10^{-4}$ &
  $6.2 \times 10^{-6}$ & $7.0 \times 10^{-6}$ \\
$\mathrm{Br}(t \to \gamma u)$ &
  $2.3 \times 10^{-6}$ & $3.0 \times 10^{-6}$ &
  $3.7 \times 10^{-6}$ & $3.6 \times 10^{-6}$ \\
$\mathrm{Br}(t \to \gamma c)$ &
  $7.7 \times 10^{-6}$ & $1.2 \times 10^{-5}$ &
  $3.7 \times 10^{-6}$ & $3.6 \times 10^{-6}$ \\
\end{tabular}
\end{center}
\caption{Best limits on top FCN couplings that can be obtained at LHC and TESLA
for one year of operation.
\label{tab:lim}}
\end{table}

We see that LHC and TESLA complement each other in the search for top FCN
vertices. The $\gamma_\mu$ couplings to the $Z$ boson can be best measured or
bound at LHC, whereas the sensitivity to the $\sigma_{\mu \nu}$ ones is better
at TESLA. For photon vertices, LHC is better for $\gamma tu$ and TESLA for
$\gamma tc$. The complementarity of LHC and TESLA also stems from the fact that
TESLA will not be able to distinguish $Ztq$ and $\gamma tq$ couplings in the
limit of its sensitivity, whereas LHC will because final states are different
and distinguish between them. On the other hand, the good charm tagging
efficiency expected at TESLA will be able to distinguish $Vtu$ and $Vtc$
couplings looking at the flavour of the final state jet, what is more difficult
to do at LHC.

\vspace{1cm}
\noindent
{\Large \bf Acknowledgements}

\vspace{0.4cm} \noindent
We thank F. del Aguila, K. M\"onig and A. Werthenbach for a cri\-ti\-cal 
rea\-ding 
of the ma\-nu\-script and S. Slabospitsky for
useful comments. JAAS thanks the members of the Theory group of DESY Zeuthen
for their warm hospitality during the realization of this
work. This work has been supported by a DAAD scholarship and
by the European Union under contract HTRN--CT--2000--00149 and by the Junta de
Andaluc\'{\i}a.

\clearpage

\end{document}